\documentclass[aps,amsmath,preprint, nofootinbib]{revtex4}

\usepackage{cancel}
\usepackage{graphicx}
\usepackage{color}
\usepackage{amsmath}
\usepackage{amsfonts}

\newcommand{\beq}{\begin{eqnarray}}
\newcommand{\eeq}{\end{eqnarray}}

\newcommand{\Slash}[1]{\ooalign{\hfil/\hfil\crcr$#1$}}

\usepackage{comment}
\usepackage{epstopdf}

\begin{document}

\title{QCD effective potential with strong $U(1)_{em}$ magnetic fields}
\author{Sho Ozaki~\footnote{
e-mail: sozaki@yonsei.ac.kr}
}

\affiliation{Institute of Physics and Applied Physics, Yonsei University, Seoul 120-749, Korea}


\begin{abstract}

We derive the analytic expression for the one-loop $SU(N_{c})$ QCD effective potential including $N_{f}$ flavor quarks which nonlinearly interact with the chromomagnetic background field and the external $U(1)_{em}$ magnetic field.
After the renormalization of couplings and fields, we obtain the correct one-loop $\beta$ functions of both QCD and QED, and the resulting effective potential satisfies the renormalization group equation.
We investigate the effect of the magnetic field on the QCD vacuum by using the effective potential, in particular for the color $SU(3)$ case with the three flavors ($u, d, s$).
Our result shows that the chromomagnetic field prefers to be parallel to the external magnetic field.
Furthermore, quark loop contributions to the effective potential with the magnetic field enhance gluonic contributions, and then the chromomagnetic condensate increases with an increasing magnetic field.
This result supports the recent observed gluonic magnetic catalysis at zero-temperature in lattice QCD.

\end{abstract}
\maketitle

\newpage

\section{Introduction}

Recently, quantum chromodynamics (QCD) which is the fundamental theory of the strong interaction, under an extremely strong magnetic field is extensively investigated.
There are numerous studies related to QCD in the presence of the strong magnetic field, including the chiral magnetic effect \cite{Kharzeev:2007jp, Fukushima:2008xe}, the magnetic catalysis \cite{Suganuma:1990nn, Gusynin:1994re, Gusynin:1994xp}, and hadron properties under the strong magnetic field \cite{Chernodub:2010qx, Hidaka:2012mz}, in terms of both effective models and lattice QCD.
The question of how the QCD vacuum and hadron properties are affected by the strong magnetic field when the strength of the magnetic field approaches or exceeds the QCD scale $eB \gtrsim \Lambda_{QCD}^{2}$ is a very interesting one.
Such a question is not academic but quite realistic.
It has been recognized that very strong magnetic fields 
are generated in the relativistic heavy ion collision. 
The strength of the magnetic field would reach the scale of $\Lambda_{QCD}^{2}$ \cite{Kharzeev:2007jp, Skokov:2009qp, Bzdak:2011yy}. 
Furthermore it is a great theoretical advantage that the lattice QCD can simulate the strongly interacting quark and gluon system in the presence of strong magnetic fields without the sign problem that appeared at finite density.

In QCD under the strong magnetic field, two kinds of strong dynamics coexist, namely, strongly interacting quark and gluon dynamics which is governed by QCD and nonlinearly interacting quark and strong magnetic field dynamics which is governed by QED. 
Since gluons do not directly interact with the magnetic field (photon)$-$only quarks do in QCD$-$the effect of the magnetic field is reflected on QCD through the quarks.
Therefore we need quarks that nonlinearly interact with electromagnetic fields as well as gluon fields.
A technique to calculate the fermion propagator nonlinearly interacting with constant electromagnetic fields and the nonlinear effective action of QED in terms of the field theory was developed by J. Schwinger \cite{Schwinger:1951nm} and is called Schwinger's proper time method.
Using the technique, several nonlinear QED effects such as the Schwinger mechanism \cite{Schwinger:1951nm},  vacuum birefringence \cite{Brezin:1971nd, BialynickaBirula:1970vy}, and photon splitting \cite{Adler:1970gg, BialynickaBirula:1970vy, Adler:1971wn} are widely explored.
Recently, the analytic expression for the vacuum polarization tensor of a photon within the one loop of a fermion including all order interaction with the external magnetic field, which is necessary for the description of vacuum birefringence, is obtained by the authors of \cite{Hattori:2012je}.
Schwinger's proper time method is also applied to evaluate the QCD effective potential \cite{Savvidy:1977as, Matinyan:1976mp, Nielsen:1978rm, Leutwyler:1980ma, Dittrich:1983ej, Elizalde:1984zv} with 
the covariantly constant background field \cite{Batalin:1976uv, Gyulassy:1986jq, Tanji:2011di}.
Although these analyses of the QCD effective potential are based on the one-loop calculation (but including all order interaction with gluon fields), they qualitatively reproduce nonperturbative features of QCD, such as the static linear potential between two opposite color charges at zero temperature \cite{Pagels:1978dd, Adler:1981as, Adler:1982rk} and the deconfinement transition at finite temperature \cite{Kapusta:1981nf, Engelhardt:1997pi, Gies:2000dw}.

In this paper, we study the QCD vacuum in the presence of the strong magnetic field at zero temperature by means of the QCD effective potential.
In order to investigate the effect of magnetic fields, we take into account the quark contributions which nonlinearly interact with the magnetic fields and the gluon fields.
Diagrammatic contributions from the quark loop to the effective potential are depicted in Fig. 1.
The calculation of these diagrams allows us to explore the sea quark effect with the magnetic field, the importance of which is emphasized in recent lattice studies \cite{Bruckmann:2013oba} in the context of the (inverse) magnetic catalysis.
Galilo and Nedelko have derived the integral form of the quark effective potential and numerically performed it \cite{Galilo:2011nh}.
In their results, the chromomagnetic(-electric) field prefers to be parallel (perpendicular) to the magnetic field.
These behaviors are confirmed by recent $SU(2)$ \cite{Ilgenfritz:2012fw, Ilgenfritz:2013ara} and $SU(3)$ \cite{Bali:2013esa} lattice QCD simulations.
Furthermore, Bali $et$ $al$. \cite{Bali:2013esa} have also reported the gluonic magnetic catalysis at zero temperature, which is an enhancement of the gluonic action density induced by the magnetic field.
In these lattice studies \cite{Ilgenfritz:2012fw, Ilgenfritz:2013ara, Bali:2013esa}, a significant correlation between the chromomagnetic component of the plaquette energy and the external magnetic field is pointed out, and then the chromomagnetic component gives the larger contribution to the plaquette energy than the chromoelectric component.
In this study, we focus on the chromomagnetic component of the gluon field and explore an influence of the external magnetic field on QCD through the quark loop contribution. 
We find that in the case of the pure chromomagnetic background, one can obtain the analytic expression for the QCD effective potential in the presence of the magnetic field.
Then, we compare our analytic results with the previous study \cite{Galilo:2011nh}, in which the proper time integral is numerically performed, and also with current lattice QCD data \cite{Ilgenfritz:2012fw, Ilgenfritz:2013ara, Bali:2013esa}.

We study the QCD effective potential with the magnetic field, in particular for the color $SU(3)$ case with the three flavors ($u, d, s$).
Our results show that the chromomagnetic field prefers to be parallel (or antiparallel) to the external magnetic field, 
which is consistent with the previous study \cite{Galilo:2011nh} and recent lattice results \cite{Ilgenfritz:2012fw, Ilgenfritz:2013ara, Bali:2013esa}.
Furthermore, including the pure Yang-Mills (YM) part (gluon and ghost loops), we also obtain the correct one-loop $\beta$ functions of both QCD and QED, and the total effective potential is renormalization-group invariant.
When the magnetic field is turned off, the QCD effective potential as a function of the chromomagnetic field has a minimum away from the origin.
This minimum corresponds to the dynamical generation of the chromomagnetic condensate \cite{Savvidy:1977as, Matinyan:1976mp, Nielsen:1978rm, Leutwyler:1980ma, Dittrich:1983ej, Elizalde:1984zv}.
When the magnetic field is turned on, the sea quark effect with the magnetic field enhances gluonic (gluon and ghost loop) contributions, and then the chromomagnetic condensate increases with an increasing magnetic field. 
This result supports the gluonic magnetic catalysis at zero temperature reported by Bali $et$ $al$.

This paper is organized as follows.
In Sec. II, we derive the quark part of the QCD effective potential with the pure chromomagnetic background and the external $U(1)_{em}$ magnetic field.
Including the YM part, we investigate the properties of the QCD effective potential in the presence of the magnetic field, especially for the color $SU(3)$ case with the three flavors ($u, d, s$) in Sec. III.
Finally, we conclude our study in Sec. IV.
In the Appendices, we provide a derivation of the one-loop effective potential for the $SU(N_{c})$ pure Yang-Mills theory, and the relation between the dimensional regularization and the cutoff regularization.

\begin{figure}
\begin{minipage}{0.9\hsize}
\begin{center}
\includegraphics[width=0.9 \textwidth]{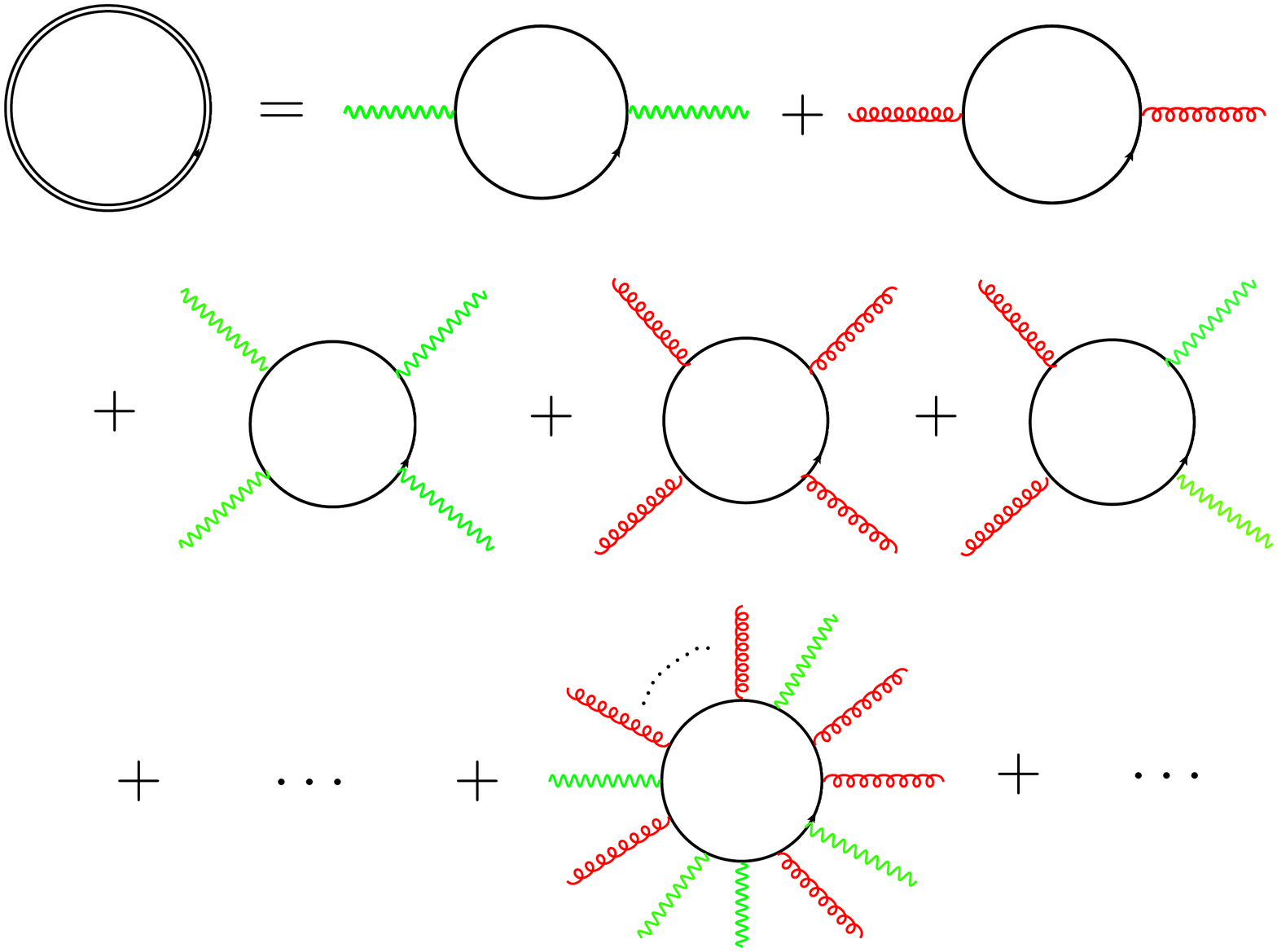}
\vskip -0.1in
\end{center}
\end{minipage}
\caption{
Quark loop nonlinearly interacting with gluon and electromagnetic fields.
}
\end{figure}

\section{One-loop effective potential of $SU(N_{c})$ QCD with pure chromomagnetic background and external $U(1)_{em}$ magnetic field}

We shall start with the $SU(N_{c})$ QCD Lagrangian with $U(1)_{em}$ electromagnetic fields which is given by
\beq
\mathcal{L} = - \frac{1}{4} F^{A}_{\mu \nu}  F^{A \mu \nu} -\frac{1}{4}f_{\mu \nu} f^{\mu \nu} + \bar{q} (i\gamma_{\mu} D^{\mu} - M_{q}) q,
\eeq
where the covariant derivative contains the photon field as well as the gluon field,
\beq
D_{\mu} = \partial_{\mu} - ig A^{A}_{\mu} T^{A} - ieQ_{q}a_{\mu},
\eeq
and the field strengths are 
\beq
F^{A}_{\mu \nu} 
&=& \partial_{\mu}A^{A}_{\nu} - \partial_{\nu}A^{A}_{\mu} + gf^{ABC}A^{B}_{\mu} A^{C}_{\nu}, \nonumber \\
f_{\mu \nu} 
&=& \partial_{\mu} a_{\nu} - \partial_{\nu} a_{\mu}.
\eeq
Here $A^{A}_{\mu}$ is the non-Abelian gauge (gluon) field, while $a^{\mu}$ is the $U(1)_{em}$ electromagnetic gauge (photon) field.
The quark field $q$ has $N_{c}$ components for the color, $N_{f}$ components for the flavor and four components for the spinor.
The quark mass and charge matrices are given by $M_{q} = {\rm{diag}} ( m_{q_{1}}, m_{q_{2}}, \cdots, m_{q_{N_{f}}})$ and $Q_{q} = {\rm{diag}}( Q_{q_{1}}, Q_{q_{2}}, \cdots, Q_{q_{N_{f}}} )$, respectively.
In this study, we consider constant external electromagnetic fields, namely, 
\beq
\partial f = 0,
\eeq
and we do not consider a quantum fluctuation of photon.
In particular, we will concentrate on constant magnetic fields later.
We apply the background field method for the non-Abelian gauge field and decompose the gauge field into a background field and a quantum fluctuation as
\beq
A^{A}_{\mu} = \hat{A}^{A}_{\mu} + \mathcal{A}^{A}_{\mu},
\eeq
where $\hat{A}^{A}_{\mu}$ is a slowly varying classical background field corresponding to the low-energy mode of the gluon field, while $\mathcal{A}^{A}_{\mu}$ is a quantum fluctuation corresponding to the high-energy mode.
We choose the covariantly constant field as a background field, which satisfies the following condition \cite{Batalin:1976uv, Gyulassy:1986jq, Tanji:2011di}
\beq
\hat{D}^{AC}_{\rho} \hat{F}^{C}_{\mu \nu} = 0,
\label{covariant_back}
\eeq
with $\hat{D}^{AC}_{\rho} = \partial_{\rho} \delta^{AC} + gf^{ABC} \hat{A}^{B}_{\rho}$ and $\hat{F}^{A}_{\mu \nu} =  \partial_{\mu} \hat{A}_{\nu}^{A} - \partial_{\nu} \hat{A}_{\nu}^{A} + g f^{ABC} \hat{A}_{\mu}^{B} \hat{A}_{\nu}^{C}$. 
Here we suppose that $\hat{F}$ is a very slowly varying field so that $\partial \hat{F} =0$ owing to the large wave length of the low-energy mode of the gluon field.
The deviation from the homogeneity of the background field is treated as the quantum fluctuation $\mathcal{A}$, as discussed in \cite{Galilo:2011nh}.
Then, $\hat{F}$ can be written as
\beq
\hat{F}^{A}_{\mu \nu} = F_{\mu \nu} \hat{n}^{A},
\eeq
where $F_{\mu \nu}$ and $\hat{n}$ are $x$ independent.
 $\hat{n}$ is a unit vector in the color space and normalized as $\hat{n}^{A} \hat{n}^{A} = 1$.
The background gauge field is proportional to $\hat{n}$,
\beq
\hat{A}^{A}_{\mu} = A_{\mu} \hat{n}^{A},
\eeq
and $F_{\mu \nu}$ has an Abelian form,
\beq
F_{\mu \nu} = \partial_{\mu} A_{\nu} - \partial_{\nu} A_{\mu}.
\eeq
Actually $\hat{A}_{\mu}^{A}$ obeys the condition (\ref{covariant_back}).
Then, the full non-Abelian field strength can be written as
\beq
F^{A}_{\mu \nu} = F_{\mu \nu} \hat{n}^{A} + \left( \hat{D}^{AC}_{\mu} \mathcal{A}^{C}_{\nu} - \hat{D}_{\nu}^{AC} \mathcal{A}_{\mu}^{C} \right) + gf^{ABC} \mathcal{A}_{\mu}^{B}
\mathcal{A}_{\nu}^{C}.
\eeq
In order to perform the functional integral of the gauge field fluctuation, we have to fix the gauge.
Here we apply the background gauge:
\beq
\hat{D}_{\mu}^{AC} \mathcal{A}^{C \mu} =0.
\eeq
We can obtain the effective action for $\hat{A}$ by performing the following functional integral
\beq
{\rm{exp}} \left[ i S_{eff}(\hat{A}) \right]
&=& \int \mathcal{D} \mathcal{A} \mathcal{D}c \mathcal{D}\bar{c} \mathcal{D} q \mathcal{D} \bar{q}
{\rm{exp}} \left\{ i \int d^{4}x \left[ - \frac{1}{4}  \left( F_{\mu \nu} \hat{n}^{A} + \left( \hat{D}^{AC}_{\mu} \mathcal{A}^{C}_{\nu} - \hat{D}_{\nu}^{AC} \mathcal{A}_{\mu}^{C} \right)
\right. \right. \right. \nonumber \\
&& \left. \left. \left. + gf^{ABC} \mathcal{A}_{\mu}^{B} \mathcal{A}_{\nu}^{C} \frac{}{} \right)^{2} 
- \frac{1}{2 \xi} \left( \hat{D}_{\mu}^{AC} \mathcal{A}^{C \mu} \right)^{2} - \bar{c}^{A} ( \hat{D}_{\mu} D^{\mu} )^{AC} c^{C} \right. \right.  \nonumber \\
&& \left. \left. + \bar{q} \left( i \gamma_{\mu} D^{\mu} - M_{q} \right)q - \frac{1}{4} f_{\mu \nu} f^{\mu \nu}
 \right] \right\},
\eeq
where $c$ is the ghost field, and $\xi$ is the gauge parameter. 
To calculate the one-loop effective action, we evaluate the functional integral for second-order field fluctuations and omit higher order terms.
From the functional integrals of the second-order gluon, ghost,  and quark field fluctuations, we get
\beq
\int \mathcal{D} \mathcal{A}  {\rm{exp}}
\left\{ \int d^{4}x \frac{-i}{2} \mathcal{A}^{A \mu} \left[
- ( \hat{D}^{2})^{AC} g_{\mu \nu} - 2 g f^{ABC} \hat{F}^{B}_{\mu \nu} 
\right] \mathcal{A}^{C \nu} \right\}
=
{\rm{det}} \left[ - ( \hat{D}^{2})^{AC} g_{\mu \nu} - 2 g f^{ABC} \hat{F}_{\mu \nu}^{B} \right]^{-1/2}, \nonumber 
\eeq
\beq
\int \mathcal{D} c \mathcal{D} \bar{c} \ {\rm{exp}}
\left\{ i \int d^{4}x \ \bar{c}^{A} \left[ - ( \hat{D}^{2} )^{AC} \right] c^{C} \right\}
&=& {\rm{det}} \left[ - ( \hat{D}^{2} )^{AC} \right]^{+1}, \nonumber \\
\int \mathcal{D} q \mathcal{D} \bar{q} \ {\rm{exp}} 
\left\{ i \int d^{4}x \  \bar{q} \left(
i \gamma_{\mu} \hat{D}^{\mu} - M_{q} \right) q \right\}
&=& {\rm{det}} \left[ i \gamma_{\mu} \hat{D}^{\mu} - M_{q} \right]^{+1}.
\label{full_actions}
\eeq
Here we take the Feynman gauge, $\xi = 1$. 
The resulting effective action for gluon and ghost parts, namely, for the YM theory is well known (see \cite{Savvidy:1977as, Matinyan:1976mp, Nielsen:1978rm, Leutwyler:1980ma, Dittrich:1983ej, Elizalde:1984zv}, and we also provide a review for a derivation of the one-loop effective action for the YM theory in Appendix A).
Integrating out the quantum fluctuation of the gluon field, we can focus on the low-energy mode of the gluon and obtain its effective action.
In the quark part, the covariant derivative includes the background field (low-energy mode) of the gluon field and the external electromagnetic field,
\beq
\hat{D}^{\mu} = \partial^{\mu} - i g \hat{A}_{\mu}^{A} T^{A} - ieQ_{q} a_{\mu}.
\eeq
In this section, we analytically derive the effective potential of the quark part in which quarks nonlinearly interact with the low-energy mode of the gluon field and the external electromagnetic field. Together with the gluon and ghost parts in (\ref{full_actions}) which contain the gluon dynamics, the effective potential allows us to investigate how the low-energy mode of the gluon field such as a gluon condensate is influenced by the external electromagnetic fields through the quark loop.

\subsection{Quark part of QCD effective potential}

The quark part of the QCD effective action is given as
\beq
i S_{q} = {\rm{log}} \ {\rm{det}} \left[ i \gamma_{\mu} \hat{D}^{\mu} - M_{q} \right].
\eeq
To calculate this log det, we have to diagonalize the matrix $\hat{n}^{A} T^{A}$.
Since this matrix is Hermitian, the diagonalization is possible. From a certain color rotation, we can get
\beq
U \hat{n}^{A} T^{A} U^{\dagger}
&=& {\rm{diag}} ( w_{1} , w_{2} , \cdots , w_{N_{c}} ).
\label{uni_matr}
\eeq
For the $SU(2)$ case, the diagonalization is straightforward, and we can easily obtain the eigenvalues $w_{1} = + 1/2$, $w_{2} = - 1/2$.
We will discuss the eigenvalues for the $SU(3)$ case in the next section.
The eigenvalues $w_{a}$ satisfy the following relations:
\beq
\sum_{a=1}^{N_{c}} w_{a}^{2} = {\rm{tr}} (\hat{n}^{A} T^{A} \hat{n}^{B} T^{B}) = \hat{n}^{A} \hat{n}^{B} {\rm{tr}} T^{A} T^{B} = \hat{n}^{A} \hat{n}^{B} \frac{1}{2}\delta^{AB} = \frac{1}{2},
\label{color_charge_prop1}
\eeq
and
\beq
\sum_{a=1}^{N_{c}} w_{a} = {\rm{tr}} (\hat{n}^{A} T^{A}) = \hat{n}^{A} {\rm{tr}} T^{A} = 0.
\label{color_charge_prop2}
\eeq
These relations play an important role in the latter results.
Using the eigenvalues $w_{a}$, the quark part of the effective action can be written as
\beq
i S_{q} = \sum_{a=1}^{N_{c}} \sum_{i=1}^{N_{f}} {\rm{log}} \left[ {\rm{det}} \left( i \Slash{D}_{a,i} - m_{q_{i}} \right) \right],
\eeq
with $D_{a,i}^{\mu} = \partial^{\mu} - i A_{a,i}^{\mu},$
where the fields $A_{a,i}^{\mu}$ are linear combinations of the gluon field $A^{\mu}$ and the photon field $a^{\mu}$ as
\beq
A_{a,i}^{\mu} = g w_{a} A^{\mu} + e Q_{q_{i}} a^{\mu}.
\eeq
Since now the gluon field $A_{\mu}$ is Abelian like a photon field $a_{\mu}$, the linear combined gauge field $A_{a,i}^{\mu}$ is also Abelian.
Therefore, the field strength of $A_{a,i}^{\mu}$ has the Abelian form 
$
F_{a,i}^{\alpha \beta} 
= \partial^{\alpha} A_{a,i}^{\beta}  - \partial^{\beta} A^{\alpha}_{a,i}. 
$
Moreover, the field strength satisfies $\partial F_{a,i}=0$.
This field strength can be expressed in terms of constant chromoelectromagnetic fields $\vec{E}_{c}$, $\vec{H}_{c}$ and $U(1)_{em}$ electromagnetic fields $\vec{E}$ and $\vec{B}$ as
\beq
F_{a,i}^{\alpha \beta}
&=& g w_{a} F^{\alpha \beta} + eQ_{q_{i}} f^{\alpha \beta} \nonumber \\
&=&
g w_{a} \left(
\begin{array}{cccc}
0 & E_{c x} & E_{c y} & E_{c z } \\
-E_{c x} & 0 & H_{c z} & - H_{c y} \\
-E_{c y} & - H_{c z} & 0 & H_{c x} \\
-E_{c z} & H_{c y} & - H_{c x} & 0 \\
\end{array}
\right)
+
e Q_{q_{i}}  \left(
\begin{array}{cccc}
0 & E_{x} & E_{y} & E_{z } \\
-E_{x} & 0 & B_{z} & - B_{y} \\
-E_{y} & - B_{z} & 0 & B_{x} \\
-E_{z} & B_{y} & - B_{x} & 0 \\
\end{array}
\right).
\label{matrix_field}
\eeq
Using the relations ${\rm{log}}[ {\rm{det}} (i \Slash{D}_{a,i} - m_{q_{i}})] = \frac{1}{2} {\rm{log}}[ {\rm{det}} ( \Slash{D}_{a,i}^{2} + m_{q_{i}}^{2} )]$ and
$
\Slash{D}^{2}_{a,i} = - D^{2}_{a,i} - \frac{1}{2} \sigma_{\alpha \beta} F^{\alpha \beta}_{a,i},
$
we proceed to evaluate the quark effective action,
\beq
i S_{q}
&=& \frac{1}{2} \sum_{a=1}^{N_{c}} \sum_{i = 1}^{N_{f}} {\rm{log}} \left[ {\rm{det}} \left( - D_{a,i}^{2} - \frac{1}{2} \sigma_{\alpha \beta} F^{\alpha \beta}_{a,i} + m_{q_{i}}^{2} \right) \right] \nonumber \\
&=& \frac{1}{2} \sum_{a=1}^{N_{c}} \sum_{i = 1}^{N_{f}} {\rm{Tr}} \left[ {\rm{log}} \left( - D_{a,i}^{2} - \frac{1}{2} \sigma_{\alpha \beta} F^{\alpha \beta}_{a,i} + m_{q_{i}}^{2} \right) 
\right] \nonumber \\
&=& \frac{1}{2} \sum_{a=1}^{N_{c}} \sum_{i = 1}^{N_{f}}  \int d^{4}x \ {\rm{tr}} \langle x |  \left[ {\rm{log}} \left( - D_{a,i}^{2} - \frac{1}{2} \sigma_{\alpha \beta} F^{\alpha \beta}_{a,i} + m_{q_{i}}^{2} \right) 
\right] | x \rangle.
\eeq
Here we use the following identity,
\beq
{\rm{log}}( \hat{M} - i \delta )
&=& \frac{1}{\epsilon} - \frac{ i^{\epsilon} }{ \epsilon \Gamma(\epsilon) } \int^{\infty}_{0} \frac{ds}{ s^{1-\epsilon} } e^{ -is( \hat{M} - i\delta) },
\label{log_id}
\eeq
in the limits of $\epsilon \to 0$ and $\delta \to 0$. The first divergent term can be omitted since this term does  not depend  on the fields. 
Furthermore, $\epsilon  \Gamma(\epsilon) = 1  + O(\epsilon)$.
Therefore, the effective action becomes
\beq
i S_{q}
&=&
 -  \frac{i^{\epsilon}}{2} \sum_{a=1}^{N_{c}} \sum_{i = 1}^{N_{f}} \int d^{4}x \int^{\infty}_{0} \frac{ds}{ s^{1 - \epsilon} } e^{-is( m_{q_{i}}^{2} - i\delta )} {\rm{tr}} 
 \langle x |  \ e^{ - i \mathcal{H}_{a,i} s }
 | x \rangle.
\eeq
Here we have defined the Hamiltonian $\mathcal{H}_{a,i}
= - D_{a,i}^{2} - \frac{ 1 }{2} \sigma_{\alpha \beta} F_{a,i}^{\alpha \beta}.
$
Now, let us consider the matrix element $\langle x^{\prime} | e^{-i \mathcal{H}_{a,i} s } | x^{\prime \prime} \rangle $.
In this matrix element, $e^{-i \mathcal{H}_{a,i} s}$ can be regarded as a time evolution operator, bringing about the time evolution of the system which is governed by the Hamiltonian $\mathcal{H}_{a,i}$.
The matrix element $\langle x^{\prime} | e^{-i \mathcal{H}_{a,i} s } | x^{\prime \prime} \rangle $ gives an amplitude for a ``particle" governed by $\mathcal{H}_{a,i}$ to travel from $x^{\prime \prime}$ to $x^{\prime}$ in a given time $s$.
Schwinger developed a technique to evaluate the matrix element in the case of the constant field of QED, called the Schwinger's proper time method~\cite{Schwinger:1951nm}.
Because in our case the combined gauge field $A_{a,i}^{\mu}$ is Abelian and satisfies $\partial F_{a,i}=0$, we can apply the same technique as in QED to calculate the matrix element $\langle x^{\prime} | e^{-i \mathcal{H}_{a,i} s } | x^{\prime \prime} \rangle $.
Then, we find 
\beq
\langle x^{\prime} |  \ e^{ - i \mathcal{H}_{a,i} s } | x^{\prime \prime} \rangle
&=& - \frac{i }{ (4 \pi)^{2} } \Psi_{a,i}(x^{\prime}, x^{\prime \prime}) e^{ - L_{a,i}(s) } s^{-2}  \nonumber \\
&&  \times {\rm{exp}} \left[ \frac{i}{4} ( x^{\prime} - x^{\prime \prime} ) F_{a,i} {\rm{coth}}(F_{a,i} s)
(x^{\prime} - x^{\prime \prime}) \right]  {\rm{exp}} \left[ \frac{i}{2} \sigma F_{a,i}s \right],
\label{matrix-element}
\eeq
where 
\beq
\Psi_{a,i}( x^{\prime}, x^{\prime \prime})
&=& {\rm{exp}} \left[ i \int^{x^{\prime}}_{x^{\prime \prime}} \left( A_{a,i}^{\mu}(x) + \frac{1}{2} F_{a,i}^{\mu \nu} (x - x^{\prime})_{\nu}\right)   dx_{\mu} \right], \nonumber \\
L_{a,i}(s)
&=& \frac{1}{2} {\rm{tr}} \ {\rm{log}} \left[ ( F_{a,i}s)^{-1} {\rm{sinh}}(F_{a,i}s) \right].
\eeq
If we replace $A^{\mu}_{a,i}$ and $F^{\mu \nu}_{a,i}$ by $ea^{\mu}$ and $ef^{\mu \nu}$, the matrix element reproduces the result of QED \cite{Schwinger:1951nm, Dittrich:1985yb, Dittrich:2000zu}.
Here the function $\Psi_{a,i}( x^{\prime}, x^{\prime \prime})$ is independent of the integration path, since the integrand $ A_{a,i}^{\mu}(x) + \frac{1}{2} F_{a,i}^{\mu \nu} (x - x^{\prime})_{\nu}$ has a vanishing curl \cite{Schwinger:1951nm, Dittrich:1985yb, Dittrich:2000zu}.
By restricting the integration path to be a straight line connecting $x^{\prime \prime}$ to $x^{\prime}$, we can simply write
\beq
\Psi_{a,i}( x^{\prime}, x^{\prime \prime})
&=& {\rm{exp}} \left[ i \int^{x^{\prime}}_{x^{\prime \prime}}  A_{a,i}^{\mu}(x)   dx_{\mu} \right].
\eeq
When we take the coincidence limit: $x^{\prime \prime} \to x^{\prime}$, this function becomes unity: $\Psi_{a,i}( x^{\prime}, x^{\prime \prime}) \to 1$.
We can also evaluate the following quantities
\beq
{\rm{tr}} \ {\rm{exp}} \left[ \frac{i}{2} \sigma F_{a,i} s \right]
&=& 4 {\rm{cos}}( a_{a,i}s ) {\rm{cosh}} (b_{a,i} s), \nonumber \\
e^{- L_{a,i}(s) }
&=& \frac{ ( a_{a,i}s )( b_{a,i}s ) }{ {\rm{sin}}( a_{a,i} s) {\rm{sinh}} ( b_{a,i}s ) },
\label{ab_factors}
\eeq
as in the QED calculation~\cite{Schwinger:1951nm, Dittrich:1985yb, Dittrich:2000zu}. 
Here $a_{a,i}$ and $b_{a,i}$ are related to the four eigenvalues $\pm F^{(1)}_{a,i}$, $\pm F^{(2)}_{a,i}$ of the field strength tensor $F_{a,i}^{\mu \nu}$ as \cite{Schwinger:1951nm}
\beq
\pm F^{(1)}_{a,i} = \pm i a_{a,i}, \ \ \ \pm F^{(2)}_{a,i} = \pm b_{a,i}
\eeq
with
\beq
a_{a,i}
 = \frac{1}{2} \sqrt{  \sqrt{ F_{a,i}^{4} + ( F_{a,i} \cdot \tilde{F}_{a,i} )^{2} } + F_{a,i}^{2} } \ , \ \ \ \ \ 
b_{a,i}
 = \frac{1}{2} \sqrt{  \sqrt{ F_{a,i}^{4} + ( F_{a,i} \cdot \tilde{F}_{a,i} )^{2} } - F_{a,i}^{2} } \ .
\eeq
The dual field strength tensor $\tilde{F}^{\mu \nu}_{a,i}$ is defined as $\tilde{F}^{\mu \nu}_{a,i}  = \frac{1}{2} \epsilon^{\mu \nu \alpha \beta} F_{a,i \alpha \beta}$.
By using (\ref{matrix_field}), $F_{a,i}^{2}$ and $F_{a,i} \cdot \tilde{F}_{a,i}$ can be expressed as
\beq
F_{a,i}^{2}
&=& 2 \left[ (g w_{a})^{2}( \vec{H}_{c}^{2} - \vec{E}_{c}^{2} ) + (eQ_{q_{i}})^{2} ( \vec{B}^{2} - \vec{E}^{2} ) + 2 g w_{a}eQ_{q_{i}}( \vec{H}_{c} \cdot \vec{B} - \vec{E}_{c} \cdot \vec{E}) \right], \nonumber  \\
F_{a,i} \cdot \tilde{F}_{a,i}
&=& -4 \left[ (g w_{a})^{2} \vec{E}_{c}\cdot \vec{H}_{c} + (eQ_{q_{i}})^{2} \vec{E}\cdot \vec{B} + g w_{a} eQ_{q_{i}} ( \vec{E}_{c} \cdot \vec{B} + \vec{E} \cdot \vec{H}_{c} ) \right].
\eeq
We see that the chromoelectromagnetic fields and $U(1)_{em}$ electromagnetic fields are coupled to each other through the quark loop.
Since $x^{\prime} \to x^{\prime \prime} = x$ in the local effective action calculation, the matrix element (\ref{matrix-element}) can be simplified by taking the trace as
\beq
 {\rm{tr}} \langle x | \ e^{- i \mathcal{H}_{a,i} s  } | x \rangle
&=&  - \frac{ i }{ 4 \pi^{2} s^{2} } \frac{ ( a_{a,i}s )( b_{a,i}s ) }{ {\rm{sin}}( a_{a,i} s ) {\rm{sinh}} ( b_{a,i} s ) } {\rm{cos}}( a_{a,i}s ) {\rm{cosh}}( b_{a,i}s ).
\eeq
The effective action of the quark part is then
\beq
iS_{q}
= \sum_{a=1}^{N_{c}} \sum_{i=1}^{N_{f}} \frac{ i^{1+\epsilon} }{ 8 \pi^{2} } \int d^{4}x \int^{\infty}_{0} \frac{ ds }{ s^{ 3 - \epsilon } }
e^{ - i s( m_{q_{i}} - i \delta ) }
\frac{ ( a_{a,i}s )( b_{a,i} s ) }{ {\rm{sin}}( a_{a,i} s) {\rm{sinh}} ( b_{a,i} s ) } {\rm{cos}}(a_{a,i} s) {\rm{cosh}} (b_{a,i} s).
\eeq 
Taking the Wick rotation for the proper time $s$,
\beq
\int^{\infty}_{0} ds = \int^{-i \infty}_{0} ds,
\eeq
and changing the integral variable $s \to - is$, the action reads
\beq
iS_{q}
= - \sum_{a=1}^{N_{c}} \sum_{i=1}^{N_{f}} \frac{i}{8\pi^{2}} \int d^{4}x \int^{\infty}_{0} \frac{ ds }{ s^{3-\epsilon} } e^{ - m_{q_{i}}^{2} s }
\frac{ (a_{a,i} s )(b_{a,i} s) }{ {\rm{sinh}}(a_{a,i} s ) {\rm{sin}}(b_{a,i} s) } {\rm{cosh}}(a_{a,i} s ) {\rm{cos}}( b_{a,i} s ).
\eeq
Here we have taken $\delta \to 0$ after the rotation.
With the effective action, the effective Lagrangian is defined as
\beq
\mathcal{L}_{q}
&=& \frac{ S_{q} }{ \int d^{4}x } \nonumber \\
&=& - \sum_{a=1}^{N_{c}} \sum_{i=1}^{N_{f}} \frac{1}{8\pi^{2}}  \int^{\infty}_{0} \frac{ ds }{ s^{3-\epsilon} } e^{ - m_{q_{i}}^{2} s }
\frac{ (a_{a,i} s)(b_{a,i} s) }{ {\rm{sinh}}(a_{a,i} s) {\rm{sin}}(b_{a,i} s) } {\rm{cosh}}(a_{a,i} s) {\rm{cos}}( b_{a,i} s ).
\eeq
This is the same form as the integral representation in \cite{Galilo:2011nh}.
Now, we consider the pure chromomagnetic background and an external magnetic field, namely, $b_{a,i} \to 0$ and
\beq
a_{a,i} = \sqrt{ \left( g w_{a} \vec{H}_{c} + e Q_{q_{i}} \vec{B} \right)^{2} }
= \sqrt{ (g w_{a} H_{c} )^{2} + (e Q_{q_{i}} B )^{2} + 2 g w_{a} e  Q_{q_{i}} H_{c} B {\rm{cos}} \theta_{H_{c}B} },
\eeq
where $H_{c} = \sqrt{ \vec{H}_{c}^{2} }$ and $B = \sqrt{ \vec{B}^{2} }$.
$\theta_{H_{c}B}$ stands for the angle between the chromomagnetic field $\vec{H}_{c}$ and the $U(1)_{em}$ magnetic field $\vec{B}$. 
Consequently, the effective Lagrangian of the quark part is given as
\beq
\mathcal{L}_{q}
&=& - \sum_{a=1}^{N_{c}} \sum_{i=1}^{N_{f}} \frac{ a_{a,i} }{ 8 \pi^{2} } \int^{\infty}_{0} \frac{ ds }{ s^{2 - \epsilon } } e^{-m_{q_{i}}^{2} s }
 \ {\rm{coth}}( a_{a,i} s).
\label{integ_Lq}
\eeq
Using the integral representation for the generalized zeta function \cite{maths},
\beq
\int^{\infty}_{0} d\tau \ \tau^{\alpha - 1} e^{-\beta \tau } {\rm{coth}}( \tau ) = \Gamma(\alpha) \left[ 2^{1-\alpha} \zeta \left( \alpha, \frac{\beta}{2} \right) - \beta^{-\alpha} \right],
\eeq
we obtain 
\beq
\mathcal{L}_{q}
&=&
\sum_{a=1}^{N_{c}} \sum_{i=1}^{N_{f}} \left\{ - \frac{ m_{q_{i}}^{4} }{ 16 \pi^{2} } \left( \frac{ 1 }{ \epsilon } - \gamma_{E} \right)
- \frac{ a_{a,i}^{2} }{ 24 \pi^{2} } \left( \frac{ 1 }{ \epsilon } - \gamma_{E} \right) \right. \nonumber \\
&& \left. + \frac{ a_{a,i}^{2} }{ 24 \pi^{2} } \left[ {\rm{log}} \left( 2 a_{a,i} \right) + 12 \zeta^{\prime} ( -1, \frac{ m_{q_{i}}^{2} }{ 2 a_{a,i} } ) - 1 \right] \right. \nonumber \\
&& \left. - \frac{ a_{a,i} m_{q_{i}}^{2} }{ 8 \pi^{2} } {\rm{ log }} \left( \frac{ 2 a_{a,i} }{ m_{q_{i}}^{2} } \right)  
+ \frac{ m_{q_{i}}^{4} }{ 16 \pi^{2} } \left[ {\rm{log}}( 2 a_{a,i} ) - 1 \right] \right\},
\label{effective_L1}
\eeq
where $\gamma_{E} = 0.577 \cdots$ is the Euler constant, and $\zeta(s,\lambda)$ is the generalized zeta function.
The derivative of the generalized zeta function $\zeta^{\prime} (s_{0}, \lambda)$ is defined as $\zeta^{\prime}(s_{0}, \lambda) = d / ds \ \zeta(s,\lambda) |_{s=s_{0}}$.
Using the relations between the dimensional regularization and the cutoff regularization (see Appendix B), we replace the divergences in (\ref{effective_L1}) by a ultraviolet cutoff $\Lambda$ as
\beq
\mathcal{L}_{q}
&=&
\sum_{a=1}^{N_{c}} \sum_{i=1}^{N_{f}} \left\{ - \frac{ 1 }{ 16 \pi^{2} } \left( \Lambda^{4} - 2m_{q_{i}}^{2} \Lambda^{2} + m_{q_{i}}^{4} {\rm{log}}  \Lambda^{2}  \right)
- \frac{ a_{a,i}^{2} }{ 24 \pi^{2} } {\rm{log}}   \Lambda^{2}  \right. \nonumber \\
&& \left. + \frac{ a_{a,i}^{2} }{ 24 \pi^{2} } \left[ {\rm{log}} \left( 2 a_{a,i}  \right) + 12 \zeta^{\prime} ( -1, \frac{ m_{q_{i}}^{2} }{ 2 a_{a,i} } ) - 1 \right] \right. \nonumber \\
&& \left. - \frac{ a_{a,i} m_{q_{i}}^{2} }{ 8 \pi^{2} } {\rm{ log }} \left( \frac{ 2 a_{a,i} }{ m_{q_{i}}^{2} } \right)  
+ \frac{ m_{q_{i}}^{4} }{ 16 \pi^{2} } \left[ {\rm{log}} \left( 2 a_{a,i}  \right) - 1 \right] \right\}.
\label{qcd_q_L}
\eeq
The first divergent terms in the bracket are independent of any fields.
Thus these terms do not contribute to the dynamics, and we can simply omit this part.
We will see soon that the second logarithmic divergent term which is proportional to $a_{a,i}^{2}$ can be absorbed by the renormalization of couplings and fields.
When $B=0$, the effective Lagrangian (\ref{qcd_q_L}) reduces to the quark part of the QCD effective Lagrangian \cite{Dittrich:1983ej, Elizalde:1984zv}, 
whereas when $H_{c}=0$ with $N_{c} = N_{f} = 1$ and $Q_{q}=1$, the Lagrangian reduces to the Euler-Heisenberg Lagrangian of QED {\cite{Heisenberg:1935qt} with pure magnetic fields \cite{Dittrich:1985yb, Dunne:2004nc}, as we expected.
In order to satisfy $\mathcal{L}_{q} \to 0 $ at the zero field point $H_{c}= B=0$, we add the following field-independent terms:
\beq
N_{c} \times \sum_{i=1}^{N_{f}} \left\{ - \frac{ m_{q_{i}}^{4} }{ 16 \pi^{2} } \left( {\rm{log}}( m_{q_{i}}^{2} ) - \frac{3}{2}  \right) \right\}.
\eeq 
Now, the effective potential of quark part $V_{q} = - \mathcal{L}_{q}$ can be written as
\beq
V_{q} = V_{q}^{fin} + V_{q}^{div},
\label{quark_loop_all}
\eeq
where
\beq
V^{fin}_{q} 
&=& \sum_{a=1}^{N_{c}} \sum_{i=1}^{N_{f}} \left\{
- \frac{ a_{a,i}^{2} }{ 24 \pi^{2} } \left[ {\rm{log}} \left( 2 a_{a,i}  \right) + 12 \zeta^{\prime} ( -1, \frac{ m_{q_{i}}^{2} }{ 2 a_{a,i} } ) - 1 \right] \right. \nonumber \\
&& \left. + \frac{ a_{a,i} m_{q_{i}}^{2} }{ 8 \pi^{2} } {\rm{ log }} \left( \frac{ 2 a_{a,i} }{ m_{q_{i}}^{2} } \right)  
- \frac{ m_{q_{i}}^{4} }{ 16 \pi^{2} } \left[ {\rm{log}} \left( \frac{ 2 a_{a,i} }{ m_{q_{i}}^{2} } \right) + \frac{1}{2} \right] \right\},
\label{quark_potential_fin}
\eeq
and
\beq
V_{q}^{div}
&=&  \sum_{a=1}^{N_{c}} \sum_{i=1}^{N_{f}}   \frac{ a_{a,i}^{2} }{ 24 \pi^{2} } {\rm{log}} \Lambda^{2} .
\label{quark_potential_div}
\eeq
In this form of the effective potential, we can take the massless limit $m_{q_{i}} \to 0$ without any infrared divergences.
In the massless limit, the finite part of the effective potential becomes
\beq
V^{fin}_{q}
&=& \sum_{a=1}^{N_{c}} \sum_{i=1}^{N_{f}} \left\{ - \frac{ a_{a,i}^{2} }{ 24\pi^{2} } \left[ {\rm{log}} ( a_{a,i} ) - c_{q} \right] \right\},
\eeq 
where $c_{q} = -12 \zeta^{\prime}(-1,0) + 1 - {\rm{log}}2 = 12 {\rm{log}}G - {\rm{log}}2$. Here we have used $\zeta^{\prime}(-1, 0) = \frac{1}{12} - {\rm{log}}G$, and $G = 1.282427 \cdots$ is the Glaisher constant.
It is worth mentioning that in the divergent part (\ref{quark_potential_div}), the cross terms $ gw_{a}eQ_{q_{i}} H_{c} B {\rm{cos}}\theta_{H_{c}B}$ in $a_{a,i}^{2}$ cancel after the summation of the color index thanks to the property of the eigenvalues $w_{a}$ (\ref{color_charge_prop2}).
Thus, the logarithmic divergences are proportional to either $H_{c}^{2}$ or $B^{2}$ as
\beq
V^{div}_{q}
&=&  \frac{N_{f}}{ 48\pi^{2} }  (gH_{c})^{2} {\rm{log}} \Lambda^{2}  +  \frac{ N_{c} }{ 24 \pi^{2} } \left(  \sum_{i=1}^{N_{f}} Q_{q_{i}}^{2}  \right) (eB)^{2} {\rm{log}} \Lambda^{2} ,
\label{potential_div}
\eeq
and then these divergent terms can be absorbed by the renormalization of couplings and fields with the tree level potential.
We will renormalize these logarithmic divergences together with that of the YM part in the next subsection.

\subsection{Total effective potential and renormalization}

The one-loop effective potential of the pure Yang-Mills theory which includes gluon and ghost parts is given by \cite{Savvidy:1977as, Matinyan:1976mp, Nielsen:1978rm, Leutwyler:1980ma, Dittrich:1983ej, Elizalde:1984zv} (see also Appendix A)
\beq
Re V_{YM} = V_{YM}^{fin} + V_{YM}^{div},
\eeq
where
\beq
V_{YM}^{fin} 
&=&  \frac{ 11 N_{c} }{ 96 \pi^{2} } (gH_{c})^{2} \left\{ {\rm{log}} \left( gH_{c}  \right) - c_{g} + \frac{1}{N_{c}} \sum_{h=1}^{N_{c}^{2}-1} v_{h}^{2} {\rm{log}} |v_{h}| \right\}, \nonumber \\
V_{YM}^{div}
&=& -\frac{ 11N_{c} }{ 96 \pi^{2} } ( gH_{c} )^{2}  {\rm{log}}  \Lambda^{2},
\eeq
and
\beq
Im V_{YM}
&=& -  \frac{N_{c}}{ 16 \pi^{2} } (gH_{c})^{2}.
\label{im_V}
\eeq
Here $c_{g} = (12 + 2 {\rm{log}}2 -12 {\rm{log}} G )/11 = 0.94556 \cdots $.
$v_{h}$ is the eigenvalue of the Hermitian matrix $(\mathcal{T}_{c})^{AC} = i f^{ABC} \hat{n}^{B}$, which can be obtained by using a certain $(N_{c}^{2}-1) \times (N_{c}^{2}-1)$ unitary matrix as
\beq
\mathcal{U} \mathcal{T}_{c} \mathcal{U}^{\dagger} = {\rm{diag}} ( v_{1}, v_{2}, \cdots, v_{N_{c}^{2}-1}).
\eeq
For the $SU(2)$ case, this diagonalization is also straightforward, and we can easily obtain the eigenvalues $v_{1} = +1$, $v_{2} = -1$ and $v_{3} = 0$.
We will discuss the eigenvalues $v_{h}$ for the $SU(3)$ case in the next section.
From the Cartan-Killing metric of $SU(N_{c})$, we can get the following property of $v_{h}$:
\beq
\sum_{h=1}^{N_{c}^{2}-1} v_{h}^{2} = {\rm{tr}} (\mathcal{T}_{c} \mathcal{T}_{c}) = \hat{n}^{A} \hat{n}^{B} (- f^{CAC^{\prime}} f^{C^{\prime}BC}) = \hat{n}^{A} \hat{n}^{B} N_{c} \delta^{AB} = N_{c}.
\label{vh_prop}
\eeq
The imaginary part of the effective potential ({\ref{im_V}}) corresponds to the Nielsen-Olesen (N-O) instability \cite{Nielsen:1978rm}.

Here we should comment on the N-O instability.
The origin of the N-O instability is well known as follows.
After the diagonalization of the gluon in the color space, the quantum fluctuations $\mathcal{A}$ correspond to the off-diagonal gluons.
The diagonal parts of $\mathcal{A}$ do not couple to the background field $\hat{A}$ and thus do not contribute to the dynamics \cite{Tanji:2011di}.
Without loss of generality, we can choose the orientation of $\vec{H}_{c}$ as the $z$ direction.
Then, the energy spectra of the massless off-diagonal gluon in the presence of the background chromomagnetic field are given by $E_{n} = \sqrt{ p_{z}^{2} + 2|gv_{h}H_{c}|( n + 1/2) - 2gv_{h}H_{c}S_{z} }$ where $n=0,1, 2 \cdots$ and $S_{z} = \pm 1$.
The lowest Landau level (LLL) $n=0$ with $S_{z}=+1$ gives the tachyonic mode for $gv_{h}H_{c} > 0$, because $E_{0} = \sqrt{ p_{z}^{2} - |gv_{h}H_{c}| }$ become pure imaginary when $p_{z}^{2} < |gv_{h}H_{c}|$.
This is the origin of the imaginary part of the effective potential, namely, the N-O instability.
Now, in the $low$ energy (strong-coupling) region, the true ground state (vacuum) of the YM theory should be stable, so the N-O instability obtained by the one-loop calculation could be stabilized by nonperturbative dynamics of the YM theory.\footnote{However in $high$-energy (weak-coupling) regions, the N-O instability (pair creation of gluons) would be a physical phenomenon
because in such energy regions the one-loop analysis should be justified.
For instance, in the early stages of relativistic heavy ion collisions, extremely strong chromomagnetic fields (as well as chromoelectric fields) are generated. This chromomagnetic field could decay into gluons through the N-O instability as discussed in \cite{Fujii:2008dd, Fujii:2009kb}.} 
Although the N-O instability in the strong-coupling region would be a nontrivial problem as well as the Schwinger mechanism in the strong-coupling region \cite{Hashimoto:2013mua},
there is one possibility to stabilize the vacuum.
In the strong-coupling region, quenched lattice QCD simulations observe the large mass of the off-diagonal gluon $M_{\mathcal{A}} \gtrsim 1$ GeV in the maximally Abelian gauge \cite{Amemiya:1998jz, Gongyo:2012jb, Gongyo:2013sha} and also in the Cho-Faddeev-Niemi decomposition \cite{Shibata:2007eq}.
Since there is no mass term in the original YM Lagrangian, it can be regarded as a dynamical generation of the off-diagonal gluon mass caused by nonperturbative gluodynamics.
Then, the large off-diagonal gluon mass could shift the energy spectrum of the LLL to $E_{0} = \sqrt{ p_{z}^{2} + M_{\mathcal{A}}^{2} - |gv_{h}H_{c}| }$ and stabilize the YM vacuum, as discussed in \cite{Kondo:2004dg, Kondo:2006ih, Kondo:2013cka}.
On the other hand, the real part of the one-loop effective potential (known as the leading log model) is qualitatively in agreement with the nonperturbative analyses for the effective potential, such as the quenched lattice QCD \cite{Ambjorn:1990wf} and the functional renormalization group \cite{Eichhorn:2010zc}.
Thus, we expect that only the real part has physical meanings in the low-energy region. 
Furthermore, the quark loop contributions (\ref{quark_loop_all}) nonlinearly interacting with chromomagnetic fields and $U(1)_{em}$ magnetic fields are always real within the one-loop level.
Therefore, in this study, we concentrate on the real part of the effective potential and investigate the effect of the magnetic field on the QCD vacuum.

Including the tree level potential and the quark part of the effective potential, the total QCD effective potential is given by
\beq
V_{eff}
&=& \frac{ H_{c}^{2} }{2} + \frac{B^{2}}{2} + V^{fin} + V^{div},
\eeq
where
\beq
V^{fin}
&=& 
+ \frac{ 11 N_{c} }{ 96 \pi^{2} } (gH_{c})^{2} \left\{ {\rm{log}} \left(  gH_{c}  \right) - c_{g} + \frac{1}{N_{c}} \sum_{h=1}^{N_{c}^{2}-1} v_{h}^{2} {\rm{log}} |v_{h}| \right\} \nonumber \\
&+&\sum_{a=1}^{N_{c}} \sum_{i=1}^{N_{f}} \left\{
- \frac{ a_{a,i}^{2} }{ 24 \pi^{2} } \left[ {\rm{log}} \left( 2 a_{a,i}  \right) + 12 \zeta^{\prime} ( -1, \frac{ m_{q_{i}}^{2} }{ 2 a_{a,i} } ) - 1 \right] \right. \nonumber \\
&& \left. + \frac{ a_{a,i} m_{q_{i}}^{2} }{ 8 \pi^{2} } {\rm{ log }} \left( \frac{ 2 a_{a,i} }{ m_{q_{i}}^{2} } \right)  
- \frac{ m_{q_{i}}^{4} }{ 16 \pi^{2} } \left[ {\rm{log}} \left( \frac{ 2 a_{a,i} }{ m_{q_{i}}^{2} } \right) + \frac{1}{2} \right] \right\},
\label{potential_fin}
\eeq
and
\beq
V^{div}
&=&  \frac{1}{2} \left\{ - \frac{1}{ (4\pi)^{2} } \left( \frac{11}{3} N_{c} - \frac{2}{3} N_{f} \right) \right\} (gH_{c})^{2} {\rm{log}}  \Lambda^{2}  \nonumber \\
&& + \frac{1}{2} \frac{ N_{c} }{ 12 \pi^{2} } \left(  \sum_{i=1}^{N_{f}} Q_{q_{i}}^{2}  \right) (eB)^{2} {\rm{log}} \Lambda^{2}.
\label{potential_div}
\eeq
Now, the divergent part of the effective potential $V^{div}$ can be absorbed by the renormalization of couplings and fields with the tree level potential.
Replacing the couplings and fields in the effective potential by bare couplings $g_{0}, e_{0}$ and bare fields $H_{c 0}, B_{0}$,
the effective potential becomes
\beq
V_{eff} = \frac{H_{c 0}^{2} }{2} + \frac{B_{0}^{2}}{2} + V_{0}^{div} + V_{0}^{fin},
\eeq
where $V_{0}^{fin}$ and $V_{0}^{div}$ are the same functions as (\ref{potential_fin}) and (\ref{potential_div}) with replacing $gH_{c}$ and $eB$ by $g_{0}H_{c0}$ and $e_{0}B_{0}$. Following a conventional renormalization procedure of the gauge theory, we rescale the couplings and fields as
\beq
g = Z_{3, QCD}^{1/2} g_{0}, \ \ \ e=Z_{3, QED}^{1/2} e_{0},
\label{rescale_coupling}
\eeq
and
\beq
H_{c 0} = Z_{3, QCD}^{1/2}H_{c}, \ \ \ B_{0} = Z_{3, QED}^{1/2} B,
\eeq
where the renormalization rescaling factors are given by
\beq
Z_{3, QCD} = 1 + \delta_{3, QCD}, \ \ \ Z_{3, QED} = 1 + \delta_{3, QED}.
\label{reno_re_factor}
\eeq
Here $\delta_{3, QCD}$ and $\delta_{3, QED}$ are counterterms.
Using these renormalized couplings $g, e$ and fields $H_{c}, B$, the effective potential can be written as
\beq
V_{eff} = \frac{1}{2} Z_{3, QCD} H^{2}_{c} + \frac{1}{2} Z_{3, QED} B^{2} + V^{div} + V^{fin},
\eeq
where $V^{fin}$ and $V^{div}$ are the same functions as (\ref{potential_fin}) and (\ref{potential_div}) with $gH_{c}$ and $eB$ because $g_{0}H_{c0} = gH_{c}$ and $e_{0}B_{0} = eB$. 
Choosing the counterterms so that the logarithmic divergences in $V^{div}$ cancel,
\beq
\delta_{3, QCD}
&=& \frac{g^{2}}{(4\pi)^{2} } \left( \frac{11}{3} N_{c} - \frac{2}{3} N_{f} \right) {\rm{log}} \left( \frac{ \Lambda^{2} }{ \mu^{2} } \right), \nonumber \\
\delta_{3, QED}
&=& - \frac{  e^{2} }{ 12\pi^{2} } N_{c} \left( \sum_{i=1}^{N_{f}} Q_{q_{i}}^{2} \right) {\rm{log}} \left( \frac{ \Lambda^{2} }{ \mu^{2} } \right),
\label{counter}
\eeq
we finally obtain the finite renormalized effective potential
\beq
V_{eff} 
&=& \frac{H_{c}^{2}}{2} + \frac{B^{2}}{2} + V_{YM} + V_{q},
\label{final_re_potential}
\eeq
with
\beq
V_{YM}
&=&  \frac{ 11 N_{c} }{ 96 \pi^{2} } (gH_{c})^{2} \left\{ {\rm{log}} \left( \frac{ gH_{c} }{ \mu^{2} } \right) - c_{g} + \frac{1}{N_{c}} \sum_{h=1}^{N_{c}^{2}-1} v_{h}^{2} {\rm{log}} |v_{h}| \right\}, \\
V_{q}
&=&\sum_{a=1}^{N_{c}} \sum_{i=1}^{N_{f}} \left\{
- \frac{ a_{a,i}^{2} }{ 24 \pi^{2} } \left[ {\rm{log}} \left( \frac{ 2 a_{a,i} }{ \mu^{2} } \right) + 12 \zeta^{\prime} ( -1, \frac{ m_{q_{i}}^{2} }{ 2 a_{a,i} } ) - 1 \right] \right. \nonumber \\
&& \left. + \frac{ a_{a,i} m_{q_{i}}^{2} }{ 8 \pi^{2} } {\rm{ log }} \left( \frac{ 2 a_{a,i} }{ m_{q_{i}}^{2} } \right)  
- \frac{ m_{q_{i}}^{4} }{ 16 \pi^{2} } \left[ {\rm{log}} \left( \frac{ 2 a_{a,i} }{ m_{q_{i}}^{2} } \right) + \frac{1}{2} \right] \right\}.
\eeq
Here we have introduced an arbitrary renormalization scale point $\mu$ in the counterterms (\ref{counter}), and thus the final expression of the effective potential explicitly contains $\mu$.
However the effective potential (\ref{final_re_potential}) should be independent of an arbitrary renormalization scale point $\mu$.
Actually, the effective potential is $\mu$-independent.
To see this, we consider the following renormalization group (RG) equation,
\beq
\mu { d \over d \mu} V (g, H_{c}, e, B, \mu )= 0,
\label{reno_g_eq1}
\eeq
where couplings and fields depend on $\mu$, namely, running couplings and fields. 
Since in this study we do not take into account the quark mass renormalization, $m_{q_{i}}$ is independent of $\mu$.
The equation (\ref{reno_g_eq1}) gives
\beq
\left[ \mu { \partial \over \partial \mu} + \beta_{QCD} { \partial \over \partial g } - 2 \gamma_{H_{c}} H_{c}^{2} { \partial \over \partial H_{c}^{2} } 
+ \beta_{QED} { \partial \over \partial e } - 2 \gamma_{B} B^{2} { \partial \over \partial B^{2} } 
 \right] V (g, H_{c}, e, B, \mu ) = 0,
 \label{ren_group_eq}
\eeq
where $\beta_{QCD}$ and $\beta_{QED}$ are $\beta$ functions of QCD and QED, defined as $\beta_{QCD} = \mu \partial g / \partial \mu$ and $\beta_{QED} = \mu \partial e / \partial \mu$. 
By using the equations (\ref{rescale_coupling}) and the renormalization rescaling factors (\ref{reno_re_factor}), we can evaluate these $\beta$ functions as
\beq
\beta_{QCD}
&=& \mu \frac{ \partial g }{ \partial \mu } = \frac{1}{2} g\mu \frac{1}{Z_{3, QCD}} \frac{ \partial Z_{3, QCD} }{  \partial \mu } = - \frac{ g^{3} }{ (4\pi)^{2} }
\left(  \frac{ 11 }{3} N_{c} -  \frac{2}{3} N_{f} \right),
\label{beta_QCD} \\
\beta_{QED}
&=& \mu \frac{ \partial e }{ \partial \mu } = \frac{1}{2} e\mu \frac{1}{Z_{3, QED}} \frac{ \partial Z_{3, QED} }{  \partial \mu } = + \frac{ e^{3} }{ 12 \pi^{2} } N_{c}  \left( \sum_{i=1}^{N_{f}} Q_{q_{i}}^{2} \right)
\label{beta_QED}.
\eeq
We obtain the correct one-loop $\beta$ functions of  both QCD and QED.
$\gamma_{H_{c}}$ and $\gamma_{B}$ are the anomalous dimensions of fields, obtained as
\beq
\gamma_{H_{c}}
&=& - { \mu \over H_{c} } { \partial H_{c} \over \partial \mu } =  {1 \over 2 } \mu  { 1 \over Z_{3, QCD } } { \partial Z_{3, QCD } \over \partial \mu } = { 1 \over g } \beta_{QCD},  \\
\gamma_{B}
&=& - { \mu \over B } { \partial B \over \partial \mu } = {1 \over 2 } \mu  { 1 \over Z_{3, QED } } { \partial Z_{3, QED } \over \partial \mu }  =  { 1 \over e } \beta_{QED}.
\label{gamma_B}
\eeq
We can easily verify that the effective potential (\ref{final_re_potential}) satisfies the RG equation (\ref{ren_group_eq}).
Accordingly, the effective potential is independent of the renormalization scale point $\mu$, provided that we take into account appropriate running couplings and fields which can be obtained by solving the differential equations (\ref{beta_QCD})$-$(\ref{gamma_B}). 

\section{Properties of color $SU(3)$ QCD effective potential with external $U(1)_{em}$ magnetic field}

\begin{figure*}[ht]
\begin{tabular}{cc}
\begin{minipage}{0.55\hsize}
\includegraphics[width=0.8 \textwidth, bb = 160 50 750 550]{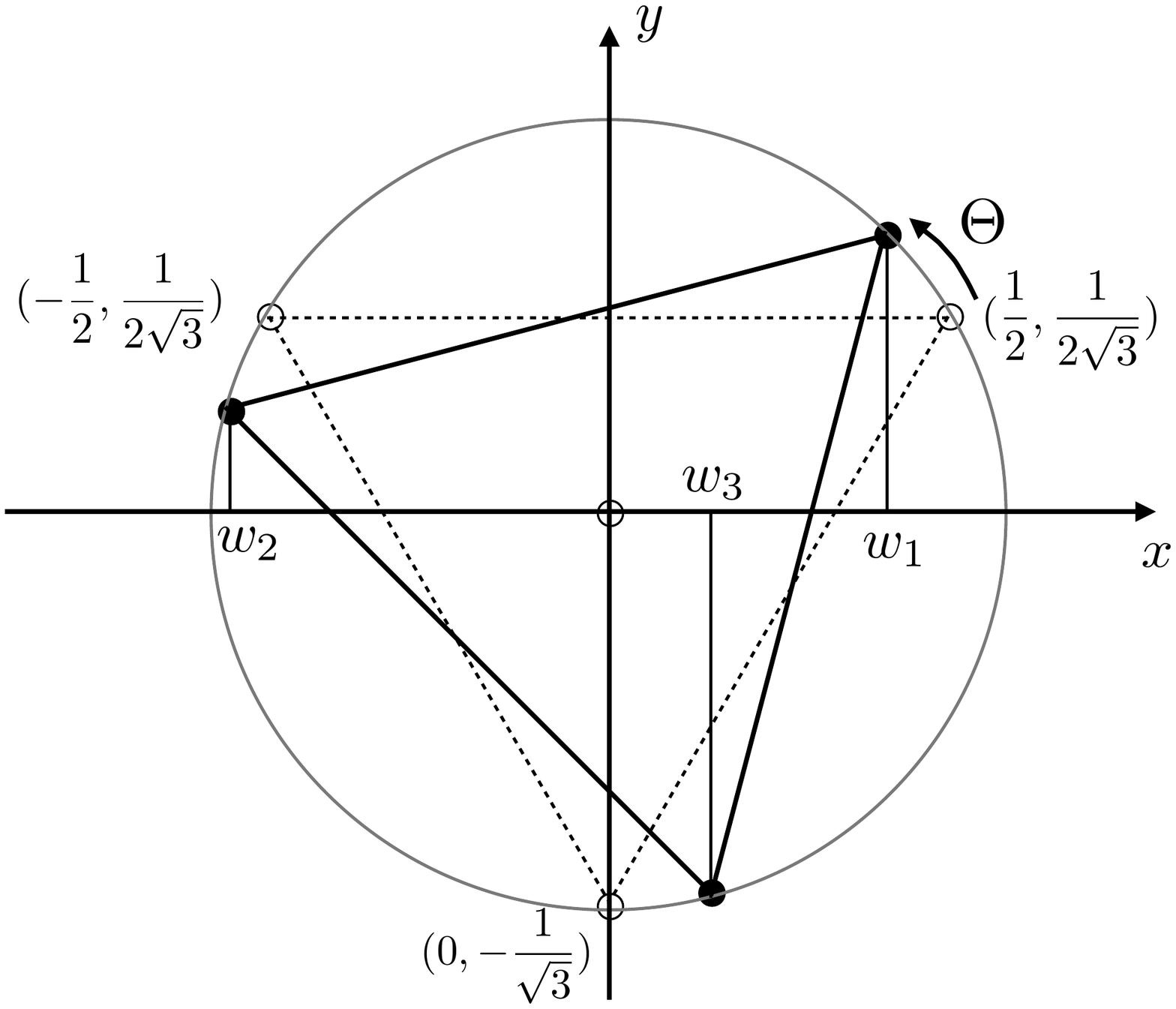}
\end{minipage}
\begin{minipage}{0.55\hsize}
\includegraphics[width=0.85 \textwidth, bb = 160 50 750 550]{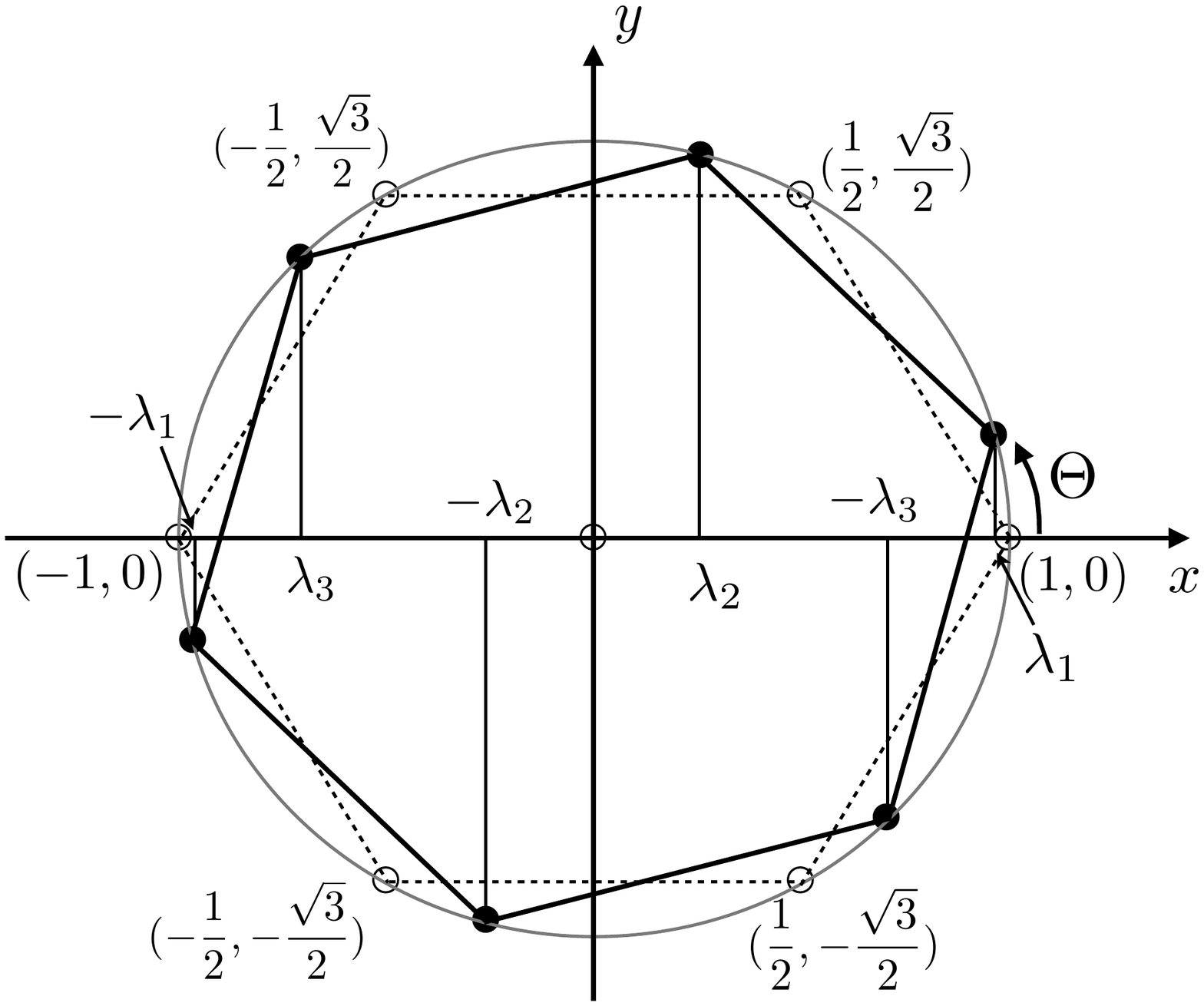}
\end{minipage}
\end{tabular}
\caption{ Diagrammatic representations of eigenvalues $w_{a}$ and $\pm \lambda_{a}$. Left panel: the rotated weight diagram. Each $w_{a}$ is given by the $x$ coordinate of  each vertex of the $\Theta$-rotated triangle.
Right panel: the rotated root diagram. Each $\pm \lambda_{a}$ is given by the $x$ coordinate of vertex of the $\Theta$-rotated hexagon. 
}
\end{figure*}

Let us now focus on the color $SU(3)$ QCD with the external $U(1)_{em}$ magnetic field.
In order to obtain the color $SU(3)$ QCD effective potential, we need the eigenvalues of $\hat{n}^{A} T^{A}$ and $\mathcal{T}_{c}$ for the color $SU(3)$ case.
First, we consider the eigenvalues of $\hat{n}^{A} T^{A}$. To get the eigenvalues, we evaluate the determinant of the $3 \times 3$ matrix $(\hat{n}^{A}T^{A} - w \mathbb{I}_{3\times3})$ and find
\beq
{\rm{det}} ( \hat{n}^{A} T^{A} - w \mathbb{I}_{3\times3} )
&=& -w^{3} + A w^{2} - B w + C \nonumber \\
&=& ( w_{1} - w ) ( w_{2} - w ) ( w_{3} - w),
\eeq
where
\beq
A = 0, \ \ \ B = - \frac{1}{4} \hat{n}^{2}, \ \ \ C = \frac{1}{12} [d_{ABC} \hat{n}^{A} \hat{n}^{B} \hat{n}^{C}].
\eeq
Therefore, the eigenvalues should satisfy the following equations:
\beq
w_{1} + w_{2} + w_{3}
&=& A, \nonumber \\
w_{1} w_{2} + w_{1} w_{3} + w_{2} w_{3}
&=& B, \nonumber \\
w_{1} w_{2} w_{3}
&=& C.
\label{omega_eqs}
\eeq
The first equation is the same as (\ref{color_charge_prop2}). We need to solve Eqs. (\ref{omega_eqs}) to get the eigenvalues.
Now, since $\hat{n}^{A} T^{A}$ is the $3 \times 3$ matrix and it is traceless, the diagonalized matrix of $\hat{n}^{A} T^{A}$ and thus the solutions of (\ref{omega_eqs}) may be expressed in terms of the two diagonal matrices $T^{3}$ and $T^{8}$ which are traceless as \cite{Tanji:2010eu}
\beq
U \hat{n}^{A} T^{A} U^{\dagger}
&=& T^{3} {\rm{cos}} \Theta - T^{8} {\rm{sin}} \Theta \nonumber \\
&=& {\rm{diag}} ( w_{1}, w_{2}, w_{3} ),
\eeq
where
\beq
w_{1} = \frac{1}{ \sqrt{3} } {\rm{cos}} \left( \Theta + \frac{ \pi }{ 6 } \right), \ \ \ w_{2} = - \frac{1}{ \sqrt{3} } {\rm{cos}} \left( \Theta - \frac{ \pi }{ 6 } \right), \ \ \
w_{3}
= \frac{1}{ \sqrt{3} } {\rm{sin}} \Theta.
\label{omegas}
\eeq 
$\Theta$ is related to the second Casimir invariant $C_{2} = [ d_{ABC} \hat{n}^{A} \hat{n}^{B} \hat{n}^{C} ]^{2}$ as
\beq
{\rm{sin}}^{2} 3 \Theta = 3 C_{2}.
\label{parametrization}
\eeq
Here we follow the notation of \cite{Tanji:2010eu}.
The relation of the notations to \cite{Nayak:2005pf, Nayak:2005yv} is discussed in \cite{Tanji:2010eu}.
The eigenvalues (\ref{omegas}) indeed satisfy the equations (\ref{omega_eqs}) and also (\ref{color_charge_prop1}).
The author of \cite{Tanji:2010eu} discussed the diagrammatic interpretation of the eigenvalues (\ref{omegas}) as depicted in the left panel of Fig. 2.
Each $w_{a}$ corresponds to the $x$ coordinate of each vertex of the rotated weight diagram.
$\Theta$ is the rotating angle of the weight diagram.

Next, we shall consider the eigenvalues of $\mathcal{T}_{c} = if^{ABC} \hat{n}^{B}$.
In order to obtain the eigenvalues, one needs to evaluate the determinant of the $8 \times 8$ matrix $ ( if^{ABC} \hat{n}^{B} - v \delta^{AC} ) $ as \cite{Nayak:2005yv}
\beq
{\rm{det}} ( if^{ABC} \hat{n}^{B} - v \delta^{AC})
&=& v^{2} ( v^{6} - A^{\prime} v^{4} + B^{\prime} v^{2} - C^{\prime} ) \nonumber \\
&=& v^{2} ( v^{2} - \lambda_{1}^{2} ) ( v^{2} - \lambda_{2}^{2} ) ( v^{2} - \lambda_{3}^{2} ),
\eeq
where
\beq
A^{\prime} = \frac{3}{2} \hat{n}^{2}, \ \ \ B^{\prime} = \frac{A^{\prime 2}}{4}, \ \ \ C^{\prime} = \frac{1}{16} [ (\hat{n}^{A}\hat{n}^{A})^{3} - 3 C_{2} ].
\eeq
There are two zero eigenvalues, and other eigenvalues $\lambda_{a}$ are all paired and satisfy
\beq
\lambda_{1}^{2} + \lambda_{2}^{2} + \lambda_{3}^{2} 
&=& A^{\prime}, \nonumber \\
\lambda_{1}^{2} \lambda_{2}^{2} + \lambda_{1}^{2} \lambda_{3}^{2} + \lambda_{2}^{2} \lambda_{3}^{2}
&=& B^{\prime}, \nonumber \\
\lambda_{1}^{2} \lambda_{2}^{2} \lambda_{3}^{2}
&=& C^{\prime}.
\label{lambdas_eq}
\eeq
The first equation of (\ref{lambdas_eq}) corresponds to the condition (\ref{vh_prop}). To get paired eigenvalues $\pm \lambda_{a}$, we need to solve these equations (\ref{lambdas_eq}).
Now, since two of the eigenvalues are zero and other eigenvalues are all paired, the number of independent variables is three.
Furthermore, the matrix $\mathcal{T}_{c}$ satisfies ${\rm{tr}} ( \mathcal{T}_{c}^{2} ) = 3$ from (\ref{vh_prop}).
Then, the diagonalized matrix of $\mathcal{T}_{c}$ and thus the solutions of (\ref{lambdas_eq}) may be expressed in terms of two diagonal matrices which satisfy (\ref{vh_prop}) as
\beq
\mathcal{U} \mathcal{T}_{c} \mathcal{U}^{\dagger}
&=& \mathcal{T}_{d}^{3} {\rm{cos}}\Theta - \mathcal{T}_{d}^{8} {\rm{sin}}\Theta,
\label{diag_Tc}
\eeq
where $\mathcal{T}_{d}^{3} = {\rm{diag}} (+1, -1, 0, +1/2, -1/2, -1/2, +1/2, 0)$ and $\mathcal{T}_{d}^{8} = {\rm{diag}} (0, 0, 0, \sqrt{3}/2, - \sqrt{3}/2, \sqrt{3}/2, - \sqrt{3}/2, 0)$ are diagonalized matrices of $if^{A3C}$ and $if^{A8C}$, respectively. 
$\mathcal{T}_{d}^{3}$ and $\mathcal{T}_{d}^{8}$ have at least two zero eigenvalues, and other eigenvalues are all paired.
From (\ref{diag_Tc}), the diagonalized matrix of $\mathcal{T}_{c}$ can be written as
\beq
\mathcal{U} \mathcal{T}_{c} \mathcal{U}^{\dagger}
&=& {\rm{diag}}( \lambda_{1}, - \lambda_{1}, 0, \lambda_{2}, - \lambda_{2}, \lambda_{3}, - \lambda_{3}, 0 ),
\eeq
where
\beq
\lambda_{1}^{2}
&=& {\rm{cos}}^{2} \Theta, \nonumber \\
\lambda_{2}^{2}
&=& {\rm{cos}}^{2} \left( \Theta + \frac{\pi}{3} \right), \nonumber\\
\lambda_{3}^{2} 
&=& {\rm{cos}}^{2} \left( \Theta + \frac{2\pi}{3} \right).
\label{lambdas}
\eeq
One can easily verify that these eigenvalues satisfy Eqs. (\ref{lambdas_eq}) when $\Theta$ obeys (\ref{parametrization}). 
These results (\ref{lambdas}) coincide with the eigenvalues obtained in \cite{Nayak:2005yv} if we use the relation of the notations mentioned in \cite{Tanji:2010eu}.
Moreover, these eigenvalues have a diagrammatic interpretation as depicted in the right panel of Fig. 2.
Each $\pm \lambda_{a}$ corresponds to the $x$ coordinate of each vertex of the rotated root diagram.
$\Theta$ is the rotating angle of the root diagram.
From the symmetries of the diagrams in Fig. 2, it is sufficient to consider the angle range $-\pi/6 \leq \Theta \leq \pi/6$.

\begin{figure}
\begin{minipage}{1.0\hsize}
\begin{center}
\includegraphics[width=1.05 \textwidth]{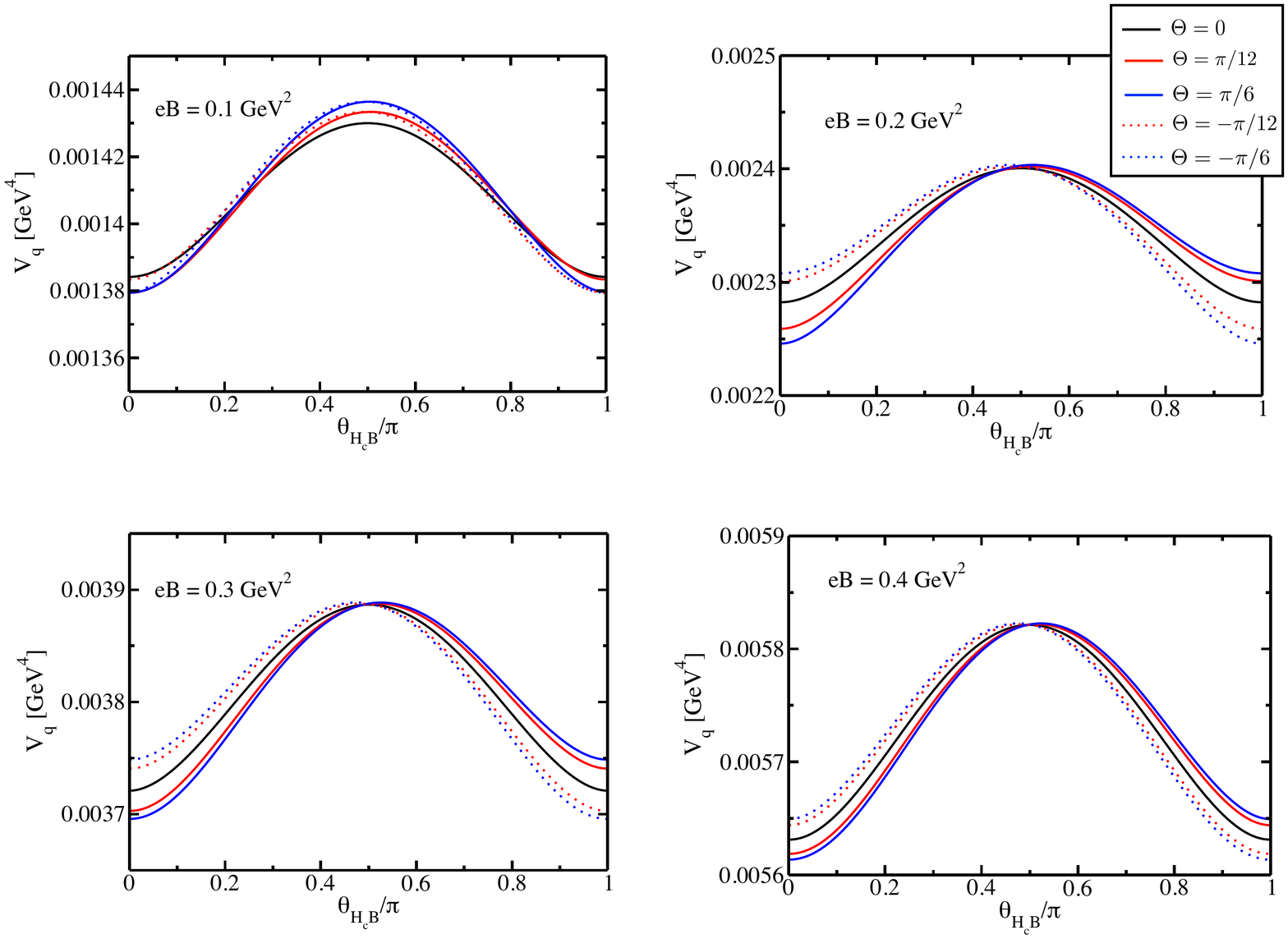}
\vskip -0.1in
\end{center}
\end{minipage}
\caption{
Quark effective potential $V_{q}$ as a function of $\theta_{H_{c}B}$ with a fixed value of the chromomagnetic field: $gH=0.2$ GeV$^{2}$ and various values of the magnetic field: $eB = 0.1,\ 0.2,\ 0.3,\ 0.4$ GeV$^{2}$ and $\Theta$: $\Theta = 0,\ \pi/12,\ \pi/6,\ -\pi/12,\ -\pi/6$.
}
\end{figure}

Now, let us investigate the properties of the color $SU(3)$ QCD effective potential with the magnetic field.
We take into account the three flavors for quarks with $Q_{u} = +2/3$ and $Q_{d} = Q_{s} = -1/3$.
The quark masses are taken as $m_{u} = m_{d} = 5$ MeV and $m_{s} = 140$ MeV.
Through this section, we take the strong and the electromagnetic coupling constants so that $\alpha_{s} = 1$ and $\alpha_{E.M.} = 1/137$ at the renormalization scale point $\mu = 1$ GeV which is chosen as a typical hadron scale. 
Since the effective potential is renormalization group invariant, the following results are $\mu$-independent, provided that we take into account appropriate running couplings and fields.

Figure 3 shows the quark part of the effective potential $V_{q}$ as a function of $\theta_{H_{c}B}$ with a fixed value of $gH_{c}$.
Here we set $gH_{c} = 0.2$ GeV$^{2}$. 
From Fig. 3, we see that $V_{q}$ is symmetric under the simultaneous transformations
$
\theta_{H_{c}B} \to - \theta_{H_{c}B} + \pi
$
and
$\Theta \to - \Theta.
$
This symmetry can be understood from the factor $a_{a,i} = \sqrt{ (g w_{a}H_{c})^{2} + (eQ_{q_{i}}B)^{2} + 2 g w_{a} eQ_{q_{i}} H_{c}B  {\rm{cos}}\theta_{H_{c}B} }$ in $V_{q}$ and the transformation law of $w_{a}$, $(w_{1}, w_{2}, w_{3}) \to (- w_{2}, - w_{1}, - w_{3})$ under $\Theta \to - \Theta$.
Another important observation in Fig. 3 is that the minima of the effective potential appear at $\theta_{H_{c}B} = 0$ (or $\pi$) with any strengths of the magnetic field. 
This means that the chromomagnetic field $\vec{H_{c}}$ prefers to be parallel (or antiparallel) to the external $\vec{B}$-field.
This result is consistent with the previous result \cite{Galilo:2011nh} in which the proper time integral is numerically performed and also with the recent lattice results \cite{Ilgenfritz:2012fw, Ilgenfritz:2013ara, Bali:2013esa}. 
In what follows, we shall take $\theta_{H_{c}B} = 0$ with $ -\frac{\pi}{6} \le \Theta \le \frac{\pi}{6}$.

\begin{figure}
\begin{minipage}{0.9\hsize}
\begin{center}
\includegraphics[width=0.7 \textwidth]{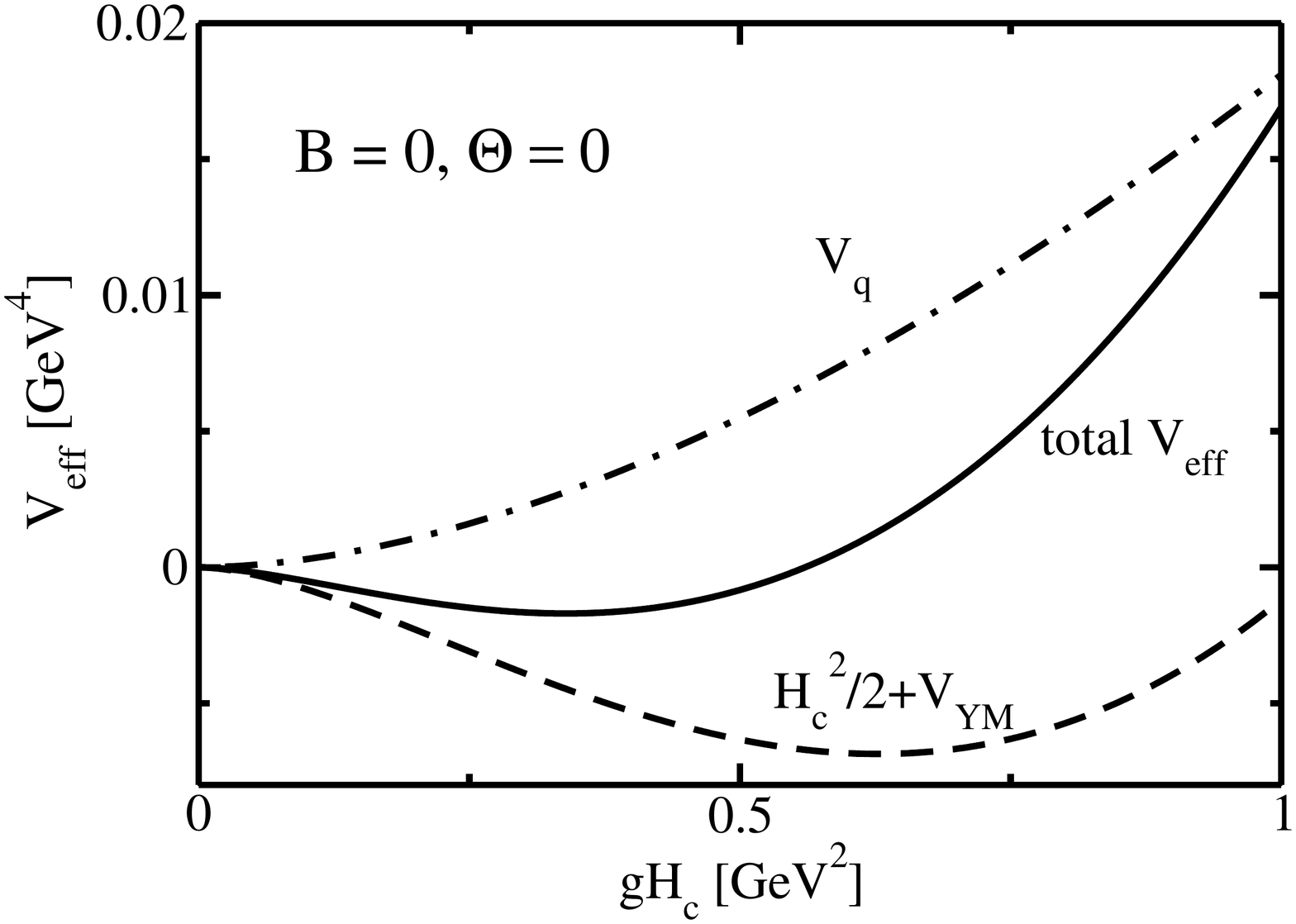}
\vskip -0.1in
\end{center}
\end{minipage}
\caption{
QCD effective potential as a function of $gH_{c}$ with $B=0$ and $\Theta=0$.
}
\end{figure}

Next we incorporate the gluon and ghost contributions into the potential.
Here we concentrate on the real part of the effective potential, as discussed in the previous section.
Figure 4 shows the QCD effective potential as a function of $gH_{c}$ with $B=0$. In this figure, we set $\Theta = 0$. 
It is well known that the one-loop YM effective potential $H_{c}^{2}/2 + V_{YM}$ has a minimum away from the origin.
This minimum corresponds to the dynamical generation of the chromomagnetic condensate \cite{Savvidy:1977as, Matinyan:1976mp, Nielsen:1978rm, Leutwyler:1980ma, Dittrich:1983ej, Elizalde:1984zv}.
As shown in Fig.~4, the quark loop contribution $V_{q}$ attenuates the gluonic contribution $V_{YM}$, owing to the opposite sign of $V_{YM}$.
Then, the minimum of the total effective potential $V_{eff}$ shifts to the left-hand side from the minimum of $H_{c}^{2}/2 + V_{YM}$.
We investigate how the chromomagnetic condensate behaves in the presence of the magnetic field.

Here, we have employed the strength of the strong coupling $\alpha_{s} = 1$ at the renormalization scale point $\mu = 1$ GeV.
Another choice of the coupling strength and the renormalization scale point will give a different position of the minimum of the effective potential.
However, the qualitative behavior of the effective potential is independent of the choice. 
In particular, the tendency of the effective potential in the presence of the magnetic field does not depend on the choice of the coupling strength and renormalization scale point.

Now we define the normalized effective potential as
\beq
\bar{V}(H_{c},B) = V_{eff}(H_{c},B) - V_{eff}(0,B),
\eeq
so that $\bar{V}(H_{c},B)$ becomes zero at $H_{c}=0$.
This normalized effective potential is also renormalization group invariant.
The second term does not affect the minimum position.
The left panel of Fig.~5 shows the magnetic field dependence of the normalized effective potential with $\Theta = 0$.
The minimum shifts to the right-hand side as the magnetic field increases.
This behavior is independent of $\Theta$.
The right panel of Fig.~5 shows the chromomagnetic condensate $(gH_{c})_{min}^{2}$ as a function of the magnetic field.
In the small magnetic field region, $(gH_{c})_{min}^{2}$ slowly increases. 
In the case of massless limit $m_{q_{i}} \to 0$, which is actually a good approximation for light quarks, we can obtain the analytic form of $(gH_{c})_{min}^{2}$ with $eB = 0$ by calculating $\partial V_{eff}/ \partial H_{c}=0$,
\beq
(gH_{c})_{min, 0}^{2} = \mu^{4} {\rm{exp}} \left\{ -\frac{ 8\pi }{ b_{0} \alpha_{s} } -1 +  \frac{2}{b_{0}} \left( \frac{11N_{c} }{ 3 }c_{g}^{\prime} - \frac{2 N_{f} }{3} c_{q}^{\prime} \right)  \right\},
\ \ \ \ \ b_{0} = \frac{ 11N_{c} }{ 3} - \frac{ 2N_{f} }{3},
\eeq
where $c_{g}^{\prime} = c_{g} - 1/N_{c} \sum_{a=1}^{N_{c} } \lambda_{a}^{2} {\rm{log}} \lambda_{a} ^{2} $ and $c_{q}^{\prime} = c_{q} - \sum_{a=1}^{N_{c}} w_{a}^{2} {\rm{log}}w_{a}^{2}$. 
Using the $(gH_{c})_{min,0}^{2}$ we find the $eB$ dependence of the chromomagnetic condensate $(gH_{c})_{min}^{2}$ for the small $eB$ region ($(gH_{c})_{min} \gg eB$) as
\beq
(gH_{c})_{min}^{2}
&=& (gH_{c})_{min,0}^{2} + \frac{(4\pi)^{2}}{b_{0}} \frac{ N_{c} }{ 12 \pi^{2} } \left( \sum_{i=1}^{N_{f}} Q_{q_{i}}^{2} \right) (eB)^{2},
\eeq
with $N_{c}=3$ and $N_{f}=3 \ (u,d,s)$ in our case.
In this expression, $(gH_{c})_{min}^{2}$ quadratically increases with respect to $eB$.
We note that the coefficient of the second term is the ratio of the coefficients of $\beta_{QCD}$ (\ref{beta_QCD}) and $\beta_{QED}$ (\ref{beta_QED}).
In the large $eB$ region, $eB > (gH_{c})_{min}$, $(gH_{c})_{min}^{2}$ still monotonically increases as the magnetic field increases.
These behaviors are quite similar to the recent lattice QCD result \cite{Bali:2013esa} with $N_{f} = 1+1+1$  staggered quarks of physical masses in which the authors insist an enhancement of the gluonic action density in the presence of the magnetic field at zero temperature, called the gluonic magnetic catalysis.



\begin{figure*}[ht]
\begin{tabular}{cc}
\begin{minipage}{0.5\hsize}
\includegraphics[width=0.8 \textwidth, bb = 140 35 745 550]{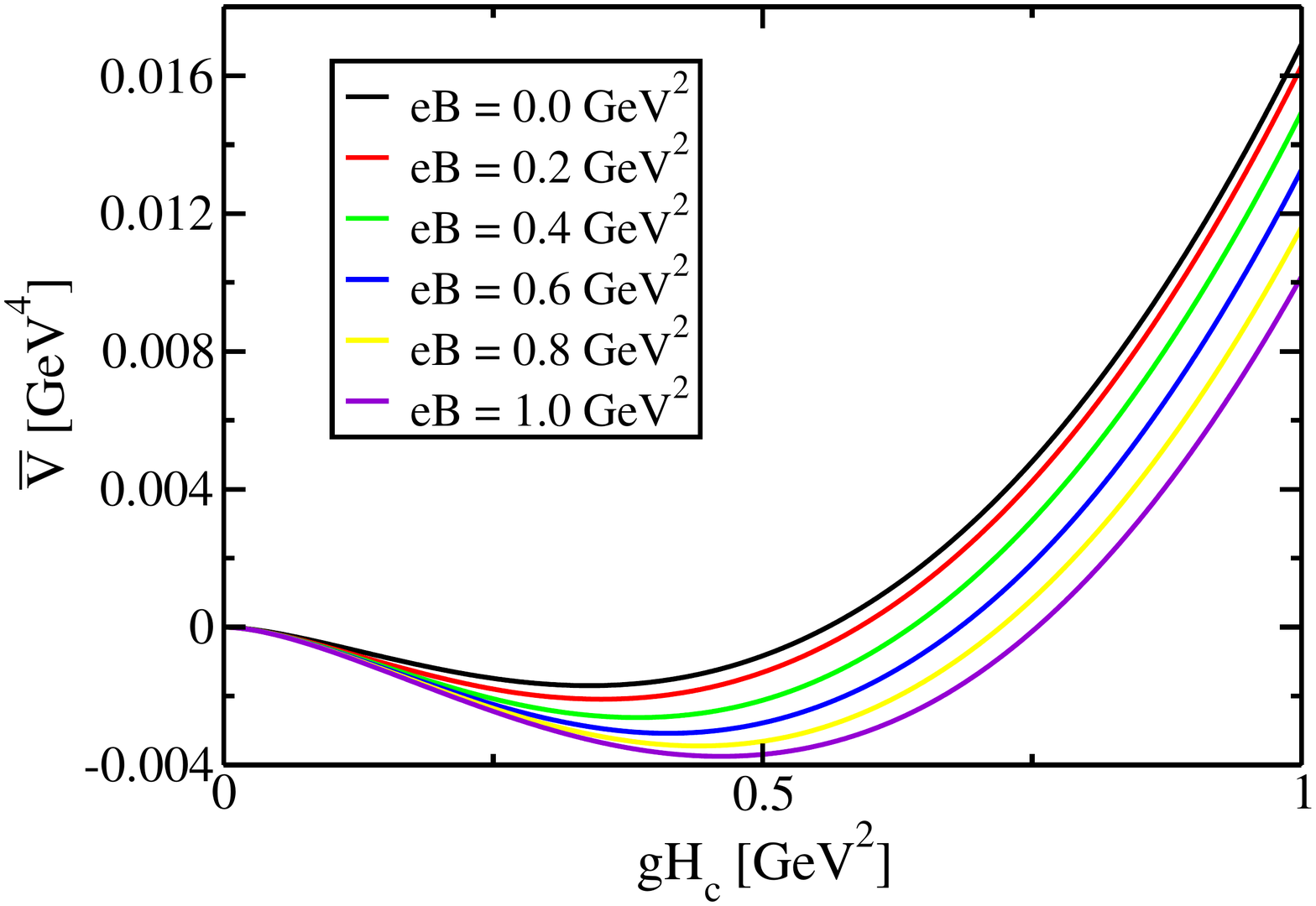}
\end{minipage}
\begin{minipage}{0.5\hsize}
\includegraphics[width=0.8 \textwidth, bb = 140 35 745 550]{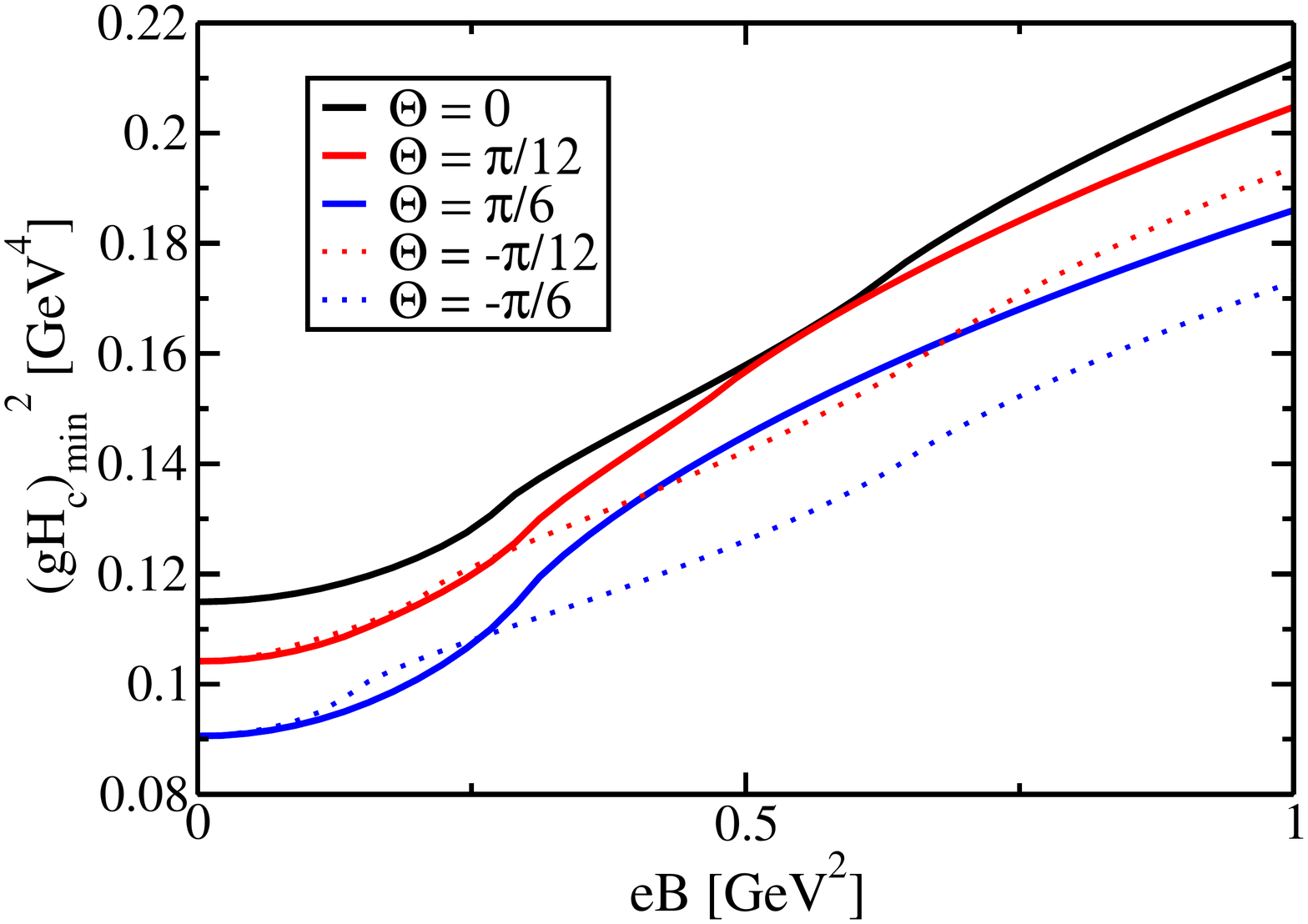}
\end{minipage}
\end{tabular}
\caption{ Left panel: the magnetic field dependence of the QCD effective potential with $\Theta =0$.
Right panel: the magnetic field dependence of the chromomagnetic condensate with various values of $\Theta$: $\Theta = 0,\ \pi/12,\ \pi/6,\ -\pi/12,\ -\pi/6$.
}
\end{figure*}

In our results, quark loop contributions which correspond to the sea quark effect discussed in \cite{Bruckmann:2013oba} should be important, since only $V_{q}$ has $B$ dependence.
To see the importance of the sea quark effect, we define the following quantity with the pure chromomagnetic background:
\beq
\Delta \bar{V}(H_{c},B)
&=& \bar{V}(H_{c},B) - \bar{V}(H_{c},0)  \nonumber \\
&=& V_{q}(H_{c},B) - V_{q}(0,B) - V_{q}(H_{c},0) \nonumber \\
&=&  \frac{i}{\int d^{4}x } {\rm{log}} \left[ \frac{ {\rm{det }} ( i \hat{\Slash{D}}(H_{c}, B) - M_{q} ) }{ {\rm{det}} ( i \hat{\Slash{D}}(H_{c}, 0 ) - M_{q} ) \ {\rm{det}}( i \hat{\Slash{D}}(0,B) - M_{q} ) } \right]. 
\label{V_change} 
\eeq
This quantity $\Delta \bar{V}$ indicates the change of the normalized effective potential $\bar{V}$ measured from the one at $B=0$, namely, the change from the uppermost (black) line to other lower (colored) lines in the left panel of Fig. 5.
Since the final line of (\ref{V_change}) contains only quark determinants, this change $\Delta \bar{V}$ is purely a sea quark effect.
A similar quantity is also calculated in the lattice study to investigate the sea quark effect with nonzero magnetic fields in terms of the reweighting technique \cite{Bruckmann:2013oba}.
Furthermore, $\Delta \bar{V}$ is renormalization group invariant since it satisfies the RG equation (\ref{ren_group_eq}).
We have numerically verified that the quantity $\Delta \bar{V}$ is always negative with any values of $gH_{c}$, $eB$ and $\Theta$
and monotonically decreasing as either $eB$ or $gH_{c}$ increases.
Figure 6 shows $\Delta \bar{V}(H,B)$ as a function of $gH_{c}$ and $eB$ in the case of $\Theta=0$ as an example.
$\Delta \bar{V}$ is negative in the whole region of $gH_{c}$-$eB$ plane and monotonically decreasing.
Now, when $B=0$, the quark loop contribution $V_{q}$ attenuates the gluonic contribution $V_{YM}$ in the total effective potential $V_{eff} = H_{c}^{2}/2 + V_{YM} + V_{q}$, owing to the opposite sign of $V_{YM}$ as we have seen in Fig. 4. From (\ref{V_change}), we can rewrite the normalized effective potential $\bar{V}(H_{c}, B)$ as
\beq
\bar{V}(H_{c}, B)
&=& \bar{V}(H_{c}, 0) + \Delta \bar{V}(H_{c}, B) \nonumber \\
&=& \frac{H_{c}^{2}}{2} + V_{YM} + \left[ V_{q}(H_{c}, 0) +  \Delta \bar{V}(H_{c},B) \right].
\eeq
Here $\Delta \bar{V}(H_{c}, B)$ can be regarded as the $B$-dependent part of the quark loop contribution, whereas $V_{q}(H_{c}, 0)$ as the $B$-independent part.
Since $\Delta \bar{V}$ is always negative and monotonically decreasing, the $B$-dependent part of the quark loop contribution enhances the gluonic contribution $V_{YM}$, which plays a completely opposite role of the $B$-independent part $V_{q}$.
Thanks to the properties of the $B$-dependent part $\Delta \bar{V}$ of the quark loop (sea quark) contribution, 
the chromomagnetic condensate $(gH_{c})_{min}^{2}$ monotonically increases with an increasing magnetic field, as we have seen in Fig. 5.
This property of the sea quark supports the gluonic magnetic catalysis at the zero temperature, observed in current lattice data \cite{Bali:2013esa}.

Although our analysis is based on the one-loop calculation but containing all order interaction with the chromomagnetic field and the external magnetic field, our results are qualitatively in agreement with recent lattice results.
Thus we expect that our analysis captures the essence of the actual physics situation.

Finally, we mention the dynamical breaking of chiral symmetry and our future works.
In this paper, we do not take into account the dynamical breaking of chiral symmetry and thus the magnetic catalysis \cite{Suganuma:1990nn, Gusynin:1994re, Gusynin:1994xp} (an enhancement of the dynamical quark mass induced by the magnetic field).
However, we expect that the magnetic catalysis also contributes to the gluonic magnetic catalysis since dynamical quark masses $M_{q}^{*}(B)$ would suppress the quark loop, especially in large $eB$ regions $eB > M_{q}^{* 2}(B=0), (gH_{c})_{min}$. 
In order to incorporate the dynamical chiral symmetry breaking and the magnetic catalysis into our framework, we should take into account a higher-order term of the quark field, namely, interaction between quarks.
In a future work, we will take the interaction between quarks such as the Nambu-Jona-Lasinio-type interaction into our effective potential and explore the effects of the dynamical chiral symmetry breaking and the magnetic catalysis.
It is also intriguing how the situation changes at finite temperature, and whether we obtain the (gluonic) $inverse$ magnetic catalysis from the sea quark effect in particular near the pseudocritical temperature $T \sim T_{c}$, as reported in \cite{Bruckmann:2013oba, Bali:2013esa}.
This interesting issue at finite temperature will also be addressed in our next project.

\begin{figure}
\begin{minipage}{0.7\hsize}
\begin{center}
\includegraphics[width=0.9 \textwidth]{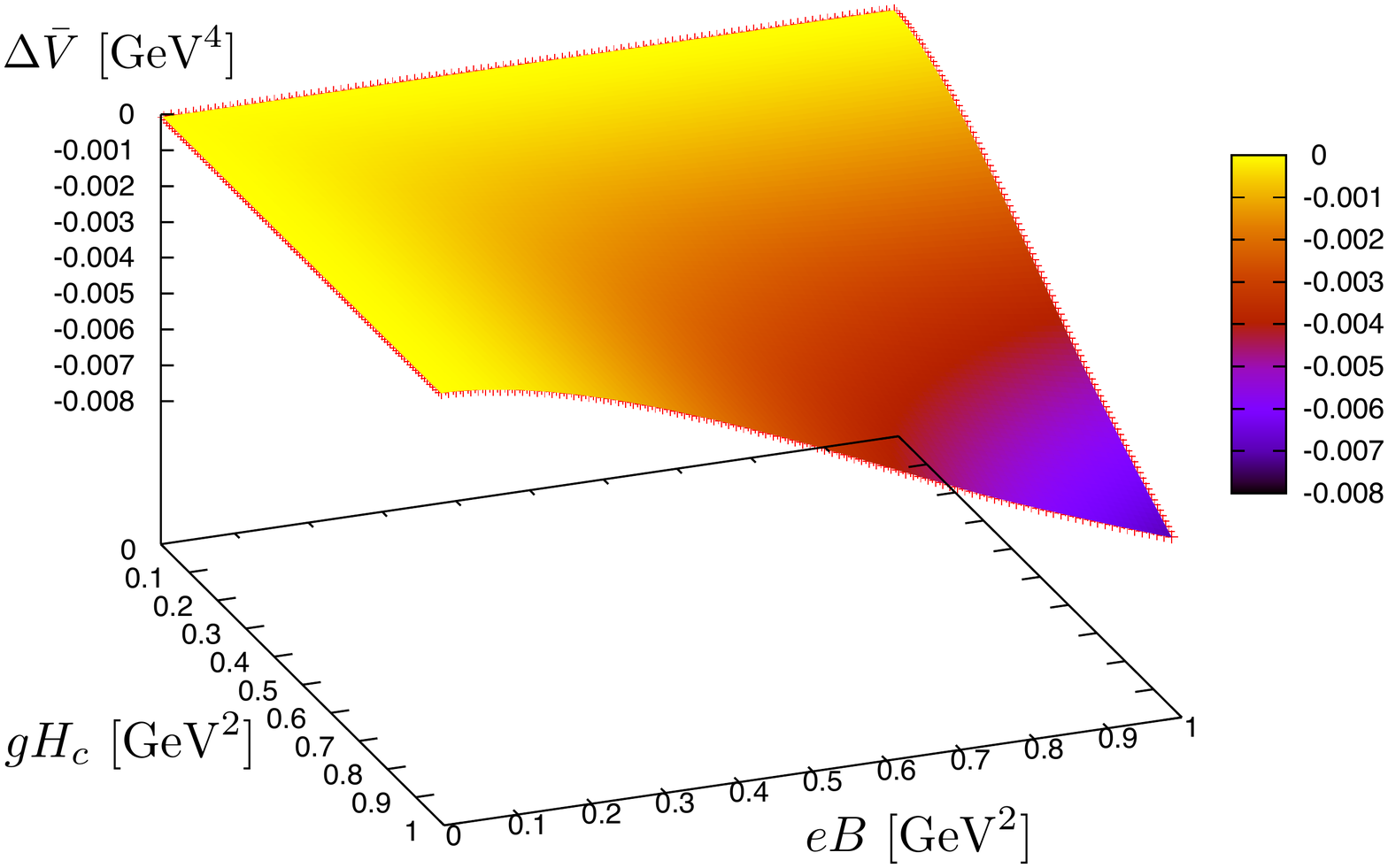}
\vskip -0.1in
\end{center}
\end{minipage}
\caption{
$\Delta \bar{V}$ as a function of $gH_{c}$ and $eB$ with $\Theta = 0$.
}
\end{figure}

\section{Summary and Conclusion}

In this paper, we derive the analytic expression for the one-loop $SU(N_{c})$ QCD effective potential including $N_{f}$ flavor quarks which nonlinearly interact with the pure chromomagnetic background field and the external $U(1)_{em}$ magnetic field.
After the renormalization of couplings and fields, we obtain the correct one-loop $\beta$ functions of both QCD and QED.
The resulting effective potential is renormalization group invariant, namely, independent of the renormalization scale point $\mu$. 
We investigate the effect of the magnetic field on the QCD effective potential in particular for the color $SU(3)$ case with the three flavors ($u,d,s$).
We find that the chromomagnetic field prefers to be parallel (or antiparallel) to the external magnetic field. This result is consistent with the previous results \cite{Galilo:2011nh} in which the proper time integral is numerically performed and also with the recent lattice results \cite{Ilgenfritz:2012fw, Ilgenfritz:2013ara, Bali:2013esa}.
Furthermore, our result shows that quark loop contributions to the effective potential (sea quark effect) with the magnetic field enhance the gluonic contributions, and thus the chromomagnetic condensate $(gH_{c})_{min}^{2}$ monotonically increases with an increasing magnetic field. 
This result supports the recent observed gluonic magnetic catalysis at zero temperature in lattice QCD \cite{Bali:2013esa}.

In a future work, we will incorporate the effects of the dynamical breaking of chiral symmetry and magnetic catalysis into the effective potential.
Furthermore, we will investigate the properties of the sea quark effect at finite temperature, especially near the pseudocritical temperature $T \sim T_{c}$.
The sea quark will play an important role for the (gluonic) $inverse$ magnetic catalysis, as recently pointed out in the lattice QCD study \cite{Bruckmann:2013oba}.

\begin{acknowledgments}
The author would like to thank S. H. Lee and K. Hattori for fruitful discussions. 
The author is also grateful to K. Itakura and S. N. Nedelko for a careful reading of the manuscript and important comments.
This work was supported by the Korean Research Foundation under Grants No. KRF-2011-0020333 and No. KRF-2011-0030621.
\end{acknowledgments}

\appendix

\section{The one-loop effective potential of $SU(N_{c})$ Yang-Mills theory with a pure chromomagnetic background}

We derive the one-loop effective potential of the $SU(N_{c})$ YM theory (see \cite{Savvidy:1977as, Matinyan:1976mp, Nielsen:1978rm, Leutwyler:1980ma, Dittrich:1983ej, Elizalde:1984zv} for the original works).
From (\ref{full_actions}), the one-loop effective action of the YM theory is given by
\beq
iS_{YM} = iS_{g} + iS_{c},
\eeq
where
\beq
iS_{g} = {\rm{log}} \ {\rm{det}} \left[ - (\hat{D}^{2})^{AC} g_{\mu \nu} - 2 g f^{ABC} \hat{F}^{B}_{\mu \nu} \right]^{-1/2},
\eeq
for the gluon part and
\beq
iS_{c} = {\rm{log}} \ {\rm{det}} \left[ - (\hat{D}^{2})^{AC} \right],
\label{ghost_p}
\eeq
for the ghost part, respectively.
Using the eigenvalues $v_{h}$ of the matrix $\mathcal{T}_{c}^{AC} = if^{ABC} \hat{n}^{B}$, 
the effective action of the gluon part becomes
\beq
iS_{g}
&=& -\frac{1}{2} \sum_{h=1}^{N_{c}^{2}-1} {\rm{log}} \ {\rm{det}} \left[ - D_{v_{h}}^{2} g_{\mu \nu} + 2 i g v_{h} F_{\mu \nu} \right] \nonumber \\
&=& -\frac{1}{2} \sum_{h=1}^{N_{c}^{2} -1} {\rm{log}} \ {\rm{det}} \left( - D_{v_{h}}^{2}  - 2  g |v_{h}| a \right) \left( - D_{v_{h}}^{2}  + 2  g |v_{h}| a \right) \nonumber \\
&& \times \left( - D_{v_{h}}^{2}  + 2  i g |v_{h}| b \right) \left( - D_{v_{h}}^{2}  - 2  ig |v_{h}| b \right),
\label{A6}
\eeq
where $D_{v_{h} \mu} = \partial_{\mu} - i g v_{h} A_{\mu}$, and 
\beq
a = \frac{1}{2} \sqrt{ \sqrt{ F^{4} + ( F \cdot \tilde{F} )^{2} } + F^{2} }, \ \ \ b = \frac{1}{2} \sqrt{ \sqrt{ F^{4} + ( F \cdot \tilde{F} )^{2} } - F^{2} } 
\eeq
are related to the eigenvalues $\pm F^{(1)}$ and $\pm F^{(2)}$ of the field strength tensor $F_{\mu \nu}$ as \cite{Schwinger:1951nm}
\beq
\pm F^{(1)} = \pm i a, \ \ \ \pm F^{(2)} = \pm b.
\eeq
Here $F^{2}$ and $F \cdot \tilde{F}$ can be expressed in terms of the chromomagnetic field $\vec{H}_{c}$ and the chromoelectric fields $\vec{E}_{c}$ as
\beq
F^{2} = 2 ( \vec{H}_{c}^{2} - \vec{E}_{c}^{2}), \ \ \ \ \ F \cdot \tilde{F} = -4 \vec{E}_{c} \cdot \vec{H}_{c}.
\eeq
The absolute values of $v_{h}$ in (\ref{A6}) appear when we explicitly calculate the eigenvalues of the matrix $- D_{v_{h}}^{2} g_{\mu \nu} + 2 i g v_{h} F_{\mu \nu} $.
We shall introduce $\rho_{v_{h}}$, which stands for any one of  $\pm 2g |v_{h}| a, \pm 2ig |v_{h}| b$, and evaluate the following action:
\beq
i S_{\rho_{v_{h}}}
&=& - \frac{1}{2} {\rm{log}} \ {\rm{det}} ( - D_{v_{h}}^{2} + \rho_{v_{h}} ) \nonumber \\
&=& - \frac{1}{2} {\rm{Tr}} \ {\rm{log}}  ( - D_{v_{h}}^{2} + \rho_{v_{h}} ) \nonumber \\
&=& - \frac{1}{2} \int d^{4}x \langle x | \ {\rm{log}} ( - D_{v_{h}}^{2} + \rho_{v_{h}} ) | x \rangle. \nonumber \\
\eeq
Using the identity (\ref{log_id}), we get
\beq
i S_{\rho_{v_{h}}}
&=& \frac{ i^{\epsilon} }{ 2 } \int d^{4}x \int^{\infty}_{0} \frac{ ds }{ s^{1 - \epsilon} } e^{ -i ( \rho_{v_{h}} - i \delta ) s } \langle x | \ e^{ - is ( - D_{v_{h}}^{2} ) } | x \rangle.
\label{S_rho}
\eeq
As mentioned in Sec. II, we can apply the Schwinger's proper time method \cite{Schwinger:1951nm} to evaluate the matrix element $\langle x^{\prime} | \ e^{ - is ( - D_{v_{h}}^{2} ) } | x^{\prime \prime} \rangle$ as in QED, since the gauge field $A_{\mu}$ is now Abelian like a photon field.
Defining the Hamiltonian $\mathcal{H}_{v_{h}} = - D_{v_{h}}^{2}$, we obtain the matrix element as
\beq
\langle x^{\prime} | \ {\rm{exp}} \left( -i \mathcal{H}_{v_{h}} s \right) | x^{\prime \prime} \rangle
&=& - \frac{i}{(4\pi)^{2}} \Psi_{v_{h}} (x^{\prime}, x^{\prime \prime}) e^{ - L_{v_{h}}(s) } s^{-2} \nonumber \\
&& \times {\rm{exp}} \left[ \frac{i}{4} ( x^{\prime} - x^{\prime \prime} ) (gv_{h}F) {\rm{coth}} ( g v_{h} F s) ( x^{\prime} - x^{\prime \prime } ) \right] ,
\eeq
where 
\beq
\Psi_{v_{h}} (x^{\prime}, x^{\prime \prime} ) 
&=& {\rm{exp}} \left[ igv_{h} \int^{x^{\prime}}_{x^{\prime \prime} } A_{\mu} dx^{\mu} \right], \nonumber \\
L_{v_{h}}(s) 
&=& \frac{1}{2} {\rm{tr}} \ {\rm{log}} \left[ (gv_{h}Fs)^{-1} {\rm{sinh}} ( g v_{h} Fs ) \right].
\eeq
In the case of the local effective action, we take $x^{\prime} \to x^{\prime \prime} = x$ and thus (\ref{S_rho}) becomes
\beq
iS_{\rho_{v_{h}}}
&=& -\frac{ i^{1+\epsilon} }{ 32 \pi^{2} } \int d^{4}x \int^{\infty}_{0} \frac{ ds }{ s^{3-\epsilon}} e^{ - L_{v_{h}} (s) } e^{ -i ( \rho_{v_{h}} - i \delta ) s }.
\eeq
Replacing $\rho_{v_{h}}$ by $\pm 2g |v_{h}| a$ and $ \pm 2ig |v_{h}| b$ and gathering all the contributions, the effective action of the gluon part can be written as
\beq
iS_{g} 
&=& \sum_{h=1}^{N_{c}^{2}-1} - \frac{ i^{1+\epsilon} }{32\pi^{2} } \int d^{4}x \int^{\infty}_{0} \frac{ds}{s^{3-\epsilon}} e^{- L_{v_{h}(s)} } \nonumber \\
&& \times \left\{ e^{-i ( -2g |v_{h}| a - i \delta ) s } + e^{ - i ( + 2g|v_{h}| a -i \delta ) s } + e^{ - i ( -2ig |v_{h}| b - i \delta) s } + e^{ - i ( +2ig |v_{h}| b - i \delta ) s} \right\}.
\eeq
We can evaluate $e^{ - L_{v_{h}} (s) }$ by using the second identity of (\ref{ab_factors}), and then the action reads
\beq
i S_{g}
&=&  \sum_{h=1}^{N_{c}^{2}-1} - \frac{ i^{1+\epsilon} }{32\pi^{2} } \int d^{4}x \int^{\infty}_{0} \frac{ds}{s^{3-\epsilon}} \frac{ ( g |v_{h}| a s )( g |v_{h}| b s) }{ {\rm{sin}}( g |v_{h}| a s ) {\rm{sinh}}( g |v_{h}| b s ) }  \nonumber \\
&& \times \left\{ e^{-i ( -2g |v_{h}| a - i \delta ) s } + e^{ - i ( + 2g |v_{h}| a -i \delta ) s } + e^{ - i ( -2ig |v_{h}| b - i \delta) s } + e^{ - i ( +2ig |v_{h}| b - i \delta ) s}
\right\} .
\eeq
Similarly, we can evaluate the effective action of the ghost part (\ref{ghost_p}).
Including the ghost part, the resulting effective action of the Yang-Mills theory is given as 
\beq
iS_{YM}
&=& \int d^{4}x - \frac{ i^{1+\epsilon} }{ 32 \pi^{2} } \sum_{h=1}^{N_{c}^{2}-1} \int^{\infty}_{0} \frac{ ds }{ s^{3-\epsilon} } e^{-\delta s }
\frac{ ( g |v_{h}| as )( g |v_{h}| bs) }{ {\rm{sin}}( g |v_{h}| as ) {\rm{sinh}}( g |v_{h}| bs ) } \nonumber \\
&& \times \left\{ e^{ +2ig |v_{h}| as } + e^{ -2ig |v_{h}| as } + e^{ +2g |v_{h}| bs } + e^{ -2g |v_{h}| bs } - 2 \right\}. 
\eeq
The last term $-2$ in the bracket corresponds to the ghost contribution.
Now, let us consider the pure chromomagnetic background, $a = \sqrt{ \vec{H_{c}}^{2} } = H_{c}$ and $b \to 0$.
Then, we get
\beq
iS_{YM}
= \int d^{4}x  - \frac{ i^{1+\epsilon} }{ 32\pi^{2} } \sum_{h=1}^{N_{c}^{2}-1} \int^{\infty}_{0} \frac{ ds }{ s^{3-\epsilon} } \frac{ g |v_{h}| H_{c}s }{ {\rm{sin}}( g |v_{h}| H_{c}s ) } 
e^{-\delta s} \left\{ e^{+2ig |v_{h}| H_{c}s } + e^{- 2ig |v_{h}| H_{c}s } \right\}.
\eeq
The effective Lagrangian can be obtain in terms of $S_{YM}$ as
\beq
\mathcal{L}_{YM}
= \frac{ S_{YM} }{  \int d^{4}x }  
= \mathcal{L}_{YM}^{stab} + \mathcal{L}_{YM}^{unstab},
\eeq
where we define the stable and unstable parts of the effective Lagrangian as \cite{Nielsen:1978rm}
\beq
\mathcal{L}_{YM}^{stab}
&=& -  i^{1+\epsilon} \sum_{h=1}^{N_{c}^{2}-1} \frac{ g |v_{h}| H_{c} }{ 16 \pi^{2} }  \int^{\infty}_{0} \frac{ ds }{ s^{2-\epsilon} } e^{-\delta s }
\left\{ \frac{ e^{-3ig |v_{h}| H_{c}s} + e^{+ig |v_{h}| H_{c}s} }{ 1 - e^{-2ig |v_{h}| H_{c}s} } - e^{+ig |v_{h}| H_{c}s} \right\}, \nonumber \\
\mathcal{L}_{YM}^{unstab}
&=& - i^{1+\epsilon} \sum_{h=1}^{N_{c}^{2}-1} \frac{ g |v_{h}| H_{c} }{ 16 \pi^{2} } \int^{\infty}_{0} \frac{ ds }{ s^{2-\epsilon} } e^{-\delta s } e^{+ig |v_{h}| H_{c}s}.
\eeq
First, we shall consider the stable part of the effective Lagrangian.
Taking the Wick rotation of the proper time $s$, the stable part reads
\beq
\mathcal{L}_{YM}^{stab}
= - \sum_{h=1}^{N_{c}^{2}-1} \frac{  g |v_{h}| H_{c}  }{ 16 \pi^{2} } i^{1+ \epsilon} \int^{-i\infty}_{0} \frac{ ds }{ s^{2-\epsilon} } e^{-\delta s } 
\left( \frac{ e^{-3ig |v_{h}| H_{c}s} + e^{ +ig |v_{h}| H_{c}s } }{ 1 - e^{-2ig |v_{h}| H_{c}s} } - e^{+ig |v_{h}| H_{c}s} \right).
\eeq
Now we change the integral variable as $s \to - is$ and take $\delta \to 0$. 
Then, the stable part becomes
\beq
\mathcal{L}_{YM}^{stab}
&=& - \sum_{h=1}^{N_{c}^{2}-1} \frac{ g |v_{h}| H_{c} }{ 16\pi^{2} } i^{1+\epsilon} \int^{\infty}_{0} \frac{ -i ds }{ ( -is )^{2-\epsilon} }
\left( \frac{ e^{-3g |v_{h}|  H_{c}s } + e^{+g |v_{h}| H_{c}s} }{ 1 - e^{-2g |v_{h}| H_{c}s} } - e^{+g |v_{h}| H_{c}s} \right) \nonumber \\
&=& \sum_{h=1}^{N_{c}^{2}-1} \frac{ g |v_{h}| H_{c} }{ 16\pi^{2} } \int^{\infty}_{0} \frac{ ds }{ s^{2-\epsilon} } 
\left( \frac{ 1 }{ {\rm{sinh}}(g |v_{h}| H_{c}s) } - e^{-g |v_{h}| H_{c}s} \right).
\label{stab_term}
\eeq
Here we consider the integral of the first term
\beq
I_{1} 
&=& \frac{ g |v_{h}| H_{c} }{ 16\pi^{2} } \int^{\infty}_{0} \frac{ ds }{ s^{2-\epsilon} } 
 \frac{ 1 }{ {\rm{sinh}}(g |v_{h}| H_{c}s) } \nonumber \\
&=&\frac{ g |v_{h}| H_{c} }{ 8\pi^{2} } \int^{\infty}_{0} \frac{ds}{s^{2-\epsilon}} \frac{ e^{-g |v_{h}| H_{c}s}  }{ 1 - e^{-2g |v_{h}| H_{c}s} }.
\eeq
Applying the following representation of the generalized zeta function \cite{maths},
\beq
\zeta(s,\lambda) = \sum_{n=0}^{\infty} \frac{ 1 }{ (n + \lambda)^{s} } = \frac{ 1 }{ \Gamma(s) } \int^{\infty}_{0} \frac{ x^{s-1} {\rm{exp}}(-\lambda x ) }{ 1 - {\rm{exp}}(-x) } dx,
\eeq
the integral $I_{1}$ can be obtained as
\beq
I_{1}
&=& \frac{ (gv_{h}H_{c})^2 }{ 4 \pi^{2} } \left\{ - \left( \frac{1}{\epsilon} - \gamma_{E} + 1 \right) \zeta(-1,1/2)  \right. \nonumber \\
&& \left. + {\rm{log}}(2g |v_{h}| H_{c}) \zeta(-1,1/2) - \zeta^{\prime}(-1,1/2) \frac{}{} \right\}.
\eeq
Next we evaluate the integral of the second term in (\ref{stab_term})
\beq
I_{2}
&=& - \frac{ g |v_{h}| H_{c} }{ 16\pi^{2} } \int^{\infty}_{0} \frac{ ds }{ s^{2-\epsilon} } e^{ - g |v_{h}| H_{c} s} \nonumber \\
&=& - \frac{ g |v_{h}| H_{c} }{ 16\pi^{2} } \frac{ \Gamma(\epsilon-1) }{ ( g |v_{h}| H_{c})^{\epsilon-1 } } \nonumber \\
&=& - \frac{ (gv_{h}H_{c})^{2} }{ 16\pi^{2} } \left\{ - \left( \frac{1}{\epsilon} - \gamma_{E} + 1 \right) + {\rm{log}}(g |v_{h}| H_{c}) \right\}.
\eeq
Now, let us consider the unstable part.
Taking the Wick rotation with the upper contour, which is a different contour from the stable part calculation, we get
\beq
\mathcal{L}_{YM}^{unstab}
&=& \sum_{h=1}^{N_{c}^{2}-1} - \frac{  g |v_{h}| H_{c} }{ 16\pi^{2} } i^{1+\epsilon}  \int^{+i\infty}_{0} \frac{ ds }{ s^{ 2-\epsilon} } e^{ig |v_{h}| H_{c}s}.
\eeq
After the Wick rotation, we have taken $\delta \to 0$. 
We change the integral variable as $s \to +is$,
and then the unstable part becomes
\beq
\mathcal{L}_{YM}^{unstab}
&=& \sum_{h=1}^{N_{c}^{2}-1} - \frac{ g |v_{h}| H_{c} }{ 16\pi^{2} } (-1)^{\epsilon} \int^{\infty}_{0} \frac{ ds }{ s^{2-\epsilon} } e^{ - g |v_{h}| H_{c}s } \nonumber \\
&=&  \sum_{h=1}^{N_{c}^{2}-1} - \frac{ (gv_{h}H_{c})^{2} }{ 16\pi^{2} } \left\{ - \left( \frac{ 1 }{\epsilon} - \gamma_{E} + 1 \right) + {\rm{log}}(g |v_{h}| H_{c}) - i\pi \right\}.
\eeq
The imaginary part appears in the last term. 
Gathering the both stable and unstable parts, we obtain the total effective Lagrangian of the YM theory as
\beq
\mathcal{L}_{YM}
&=& \sum_{h=1}^{N_{c}^{2}-1} \left\{ \frac{ (gv_{h}H_{c})^{2} }{4\pi^{2} } \left[ - \left( \frac{1}{\epsilon} - \gamma_{E} + 1 \right) \zeta(-1,1/2) + {\rm{log}}( 2g |v_{h}| H_{c}) \zeta(-1,1/2) - \zeta^{\prime}(-1,1/2) \right] \right.  \nonumber \\
&& \left. - \frac{ (gv_{h}H_{c})^{2} }{ 8\pi^{2} } \left[ - \left( \frac{1}{\epsilon} - \gamma_{E} + 1 \right) + {\rm{log}}(g |v_{h}| H_{c}) \right] \right\} \nonumber \\
&& + i \sum_{h=1}^{N_{c}^{2}-1} \frac{ ( g v_{h}H_{c})^{2} }{ 16 \pi }.
\eeq
Using (\ref{vh_prop}), the imaginary part of the Lagrangian can be written as
\beq
Im \mathcal{L}_{YM}
&=& \frac{ N_{c} }{ 16 \pi } (gH_{c})^{2}.
\eeq 
With $\zeta(-1, 1/2) = \frac{1}{24}$ and  $\zeta^{\prime}(-1,1/2) = - \frac{ 1 }{ 24 } {\rm{log}}2 - \frac{1}{24} + \frac{1}{2} {\rm{log}}G$,
the real part reads
\beq
Re \mathcal{L}_{YM}
&=& \frac{ 11N_{c} }{ 96 \pi^{2} } (gH_{c})^{2} \left( \frac{1}{\epsilon} - \gamma_{E} \right)  \nonumber \\
&& -  \frac{ 11N_{c} }{ 96 \pi^{2} } (gH_{c})^{2} \left\{ {\rm{log}}(gH_{c})  - c_{g}  
 + \frac{1}{N_{c}} \sum_{h=1}^{N_{c}^{2}-1} v_{h}^{2} {\rm{log}} |v_{h}| \right\},
\eeq
where $c_{g} = ( 12 + 2 {\rm{log}}2 - 12 {\rm{log}}G ) /11 = 0.94556 \cdots$.
Therefore, the total effective potential of the Yang-Mills theory $V_{YM} = - \mathcal{L}_{YM}$ is given as
\beq
V_{YM}
&=&  Re V_{YM} + Im V_{YM},
\eeq
where the real part is
\beq
Re V_{YM} 
&=& V^{fin}_{YM} + V_{YM}^{div},
\eeq
with
\beq
V_{YM}^{fin}
&=& + \frac{ 11 N_{c} }{ 96 \pi^{2} } (gH_{c})^{2} \left\{ {\rm{log}}(gH_{c}) - c_{g} + \frac{1}{N_{c}} \sum_{h=1}^{N_{c}^{2}-1} v_{h}^{2} {\rm{log}} |v_{h}| \right\},  \nonumber \\
V_{YM}^{div}
&=& - \frac{ 11 N_{c} }{ 96 \pi^{2} } (gH_{c})^{2} \left( \frac{ 1 }{ \epsilon} - \gamma_{E} \right),
\eeq
and
the imaginary part is
\beq
Im V_{YM} = - \frac{ N_{c} }{ 16 \pi } (gH_{c})^{2}.
\eeq
The resulting one-loop effective potential of the $SU(N_{c})$ YM theory coincides with the result of \cite{Dittrich:1983ej, Elizalde:1984zv}.


\section{Relation between dimensional regularization and cutoff regularization}

Considering the week field expansion of the (\ref{integ_Lq}), we obtain
\beq
\mathcal{L}_{q}
&=& - \sum_{a=1}^{N_{c}} \sum_{i=1}^{N_{f}} \frac{ a_{a,i} }{ 8 \pi^{2} } \int^{\infty}_{0} \frac{ ds }{ s^{ 2 - \epsilon} } e^{ - m_{q_{i}}^{2} s} {\rm{coth}}( a_{a,i}s ) \nonumber \\
&\to& - \sum_{a=1}^{N_{c}} \sum_{i=1}^{N_{f}} \frac{1}{8\pi^{2}} \int^{\infty}_{0} \frac{ ds }{ s^{3} } e^{- m_{q_{i}}^{2} s } \left( 1 + \frac{1}{3} ( a_{a,i}s )^{2} \right).
\label{divergents}
\eeq
We expect that these divergent integrals in the second line correspond to the divergent terms in (\ref{effective_L1}).
In order to estimate the divergences, we consider the following integrals:
\beq
\mathcal{I}_{1} = \int_{0}^{\infty} \frac{ ds }{ s^{3} } e^{-m^{2}s}, \ \ \ \ \mathcal{I}_{2} = \int^{\infty}_{0} \frac{ ds }{ s} e^{-m^{2}s}.
\eeq
First, we calculate the $\mathcal{I}_{2}$ by using the dimensional regularization and the cutoff regularization, and we get
\beq
\mathcal{I}_{2}^{(d)} = \int^{\infty}_{0} \frac{ ds }{ s^{1 - \epsilon } } e^{-m^{2}s} = \frac{1}{ \epsilon} - \gamma_{E} - {\rm{log}}m^{2} + O(\epsilon), \nonumber \\
\mathcal{I}_{2}^{(c)} = \int^{\infty}_{1/\Lambda^{2}} \frac{ ds }{ s } e^{-m^{2}s} = {\rm{log}} \Lambda^{2} -  {\rm{log}}m^{2} + O( \frac{ m^{2} }{ \Lambda^{2} }).
\eeq
Since the ${\rm{log}}m^{2}$ term is common, we find the relation between dimensional regularization and the cutoff regularization as
\beq
\frac{ 1 }{ \epsilon } - \gamma_{E} \leftrightarrow {\rm{log}} \Lambda^{2}.
\eeq
We have used the relation in the replacement of (\ref{effective_L1}) by (\ref{qcd_q_L}).
We note the ${\rm{log}}m^{2}$ term does not appear in the results of (\ref{effective_L1}) and (\ref{qcd_q_L}). Therefore we can take the massless limit in (\ref{qcd_q_L}) without any infrared divergences.

Next we consider the $\mathcal{I}_{1}$. We evaluate the $\mathcal{I}_{1}$ by using the dimensional regularization and the cutoff regularization, and the results are given as
\beq
\mathcal{I}_{1}^{(d)} 
&=&  \int^{\infty}_{0} \frac{ ds }{ s^{3 - \epsilon } } e^{-m^{2}s} =  \frac{ m^{4} }{2} \left( \frac{ 1 } { \epsilon} - \gamma_{E} \right) + \frac{3}{4} m^{4} - \frac{ m^{4} }{2} {\rm{log} m^{2} + O(\epsilon}), \nonumber \\
\mathcal{I}_{1}^{(c)} 
&=& \int^{\infty}_{1/\Lambda^{2} } \frac{ ds }{ s^{3} } e^{-m^{2}s} 
= \frac{1}{2} \left( \Lambda^{4} - 2 m^{2} \Lambda^{2} + m^{4} {\rm{log}} \Lambda^{2} \right) + \frac{3}{4} m^{4} - \frac{ m^{4} }{2} {\rm{log}} m^{2} + O( \frac{m^{2}}{\Lambda^{2}}  ).
\eeq
Therefore, we read off the following relation:
\beq
m^{4} \left( \frac{1}{\epsilon} - \gamma_{E} \right) \leftrightarrow  \left( \Lambda^{4} - 2 m^{2} \Lambda^{2} + m^{4} {\rm{log}} \Lambda^{2} \right).
\eeq
We have also used the relation in the replacement of (\ref{effective_L1}) by (\ref{qcd_q_L}). However, these divergent terms do not depend on any fields in (\ref{qcd_q_L}), so we have omitted these divergences.

\end{document}